\documentclass
[%
superscriptaddress,
longbibliography,
 amsmath,amssymb,
aps,
10.5pt]{revtex4-2}
\usepackage{mathrsfs}
\usepackage{siunitx}
\usepackage[dvips]{epsfig}
\usepackage{amsfonts}
\usepackage{xcolor}
\def\bm#1{\mbox{\boldmath{$#1$}}}
\def\rr#1{(\ref{#1})}

\newcommand{\be}{\begin{equation}}
\newcommand{\ee}{\end{equation}}
\usepackage{graphicx}
\usepackage{bm}

\graphicspath{ {./Chiral_Temperature_Figures} }

\begin{document}

\preprint{APS/123-QED}

\title{Oscillating electroosmotic flow in channels and capillaries with modulated wall charge distribution}
\author{A. Shrestha}
\affiliation{Center for Computation and Theory of Soft Materials, Northwestern University, Evanston IL 60208 USA
}
\affiliation{Department of Physics and Astronomy, Northwestern University, Evanston, IL 60208 USA}

\author{E. Kirkinis}
\affiliation{Center for Computation and Theory of Soft Materials, Northwestern University, Evanston IL 60208 USA
}
\affiliation{Department of Materials Science \& Engineering, Robert R. McCormick School of Engineering and Applied Science, Northwestern University, Evanston IL 60208 USA}

\author{M. Olvera de la Cruz}
\affiliation{Center for Computation and Theory of Soft Materials, Northwestern University, Evanston IL 60208 USA
}
\affiliation{Department of Physics and Astronomy, Northwestern University, Evanston, IL 60208 USA}
\affiliation{Department of Materials Science \& Engineering, Robert R. McCormick School of Engineering and Applied Science, Northwestern University, Evanston IL 60208 USA}

\date{\today}

\begin{abstract}
Electrolyte-filled channels with modulated wall charge distribution subjected to an applied DC electric field, form time-independent vortices whose sense of circulation is determined by the field direction [Physical Review Letters $ \mathbf{75}, 755, (1995)$]. In this paper we show that an electrolyte in a channel or cylindrical capillary subjected to an external \emph{alternating} (AC) electric field gives rise to various laminar flow structures, including vortices whose sense of circulation changes with the period of oscillation of the applied AC field. The introduction of a period
of oscillation lifts certain degeneracies associated with its time-independent counterpart. 
Although, in general, the mass flux vanishes, the charge flux is nonzero. 
The flow is accompanied by a longitudinal (oscillating) advective current that displays hysteresis accompanied by a diverging and negative self-similar conductance that depends on the applied voltage [Nano Letters $\mathbf{10}, 2674, (2010)$]. We show that this behavior can be interpreted with respect to a ``memory retention time'', that depends on frequency, viscosity and the Debye length and could thus form the impetus for investigating control protocols of signal carriers. 
\end{abstract}

\maketitle

\section{Introduction}

Microfluidic devices employ static channel networks to continuously activate liquids via time-independent drives. Such flows can be mediated by applied pressure gradients or electroosmosis in channels filled with ionic solutions. These static devices have been constructed using various methods such as lithography, micromilling, and injection moulding \cite{Stone2004,Squires2005,Erickson2004}. However, such predefined fabrications limit their functionalities in terms of applications that require an adjustable control of the fluid flow state. On the other hand, configurable microfluidic devices that are physically actuated by means of device assembly can be modified from a base state to a specified functional state. This has been achieved for example, through modular assembly, pattering hydrophobic surfaces, and paraffin structuring \cite{Paratore2022,Stroock2003}. Attaining regulated fluid flows in channels and capillaries that go beyond a fixed number of states is crucial for many microfluidic applications such as targeted solute delivery, mixing-induced chemical reactions, and volume-specified fluid transport.

Most studies of electrolyte transport assume that the channel walls are uniformly charged. However, acquired through fabrication or other routes, charge adsorption may change the surface charge distribution, for instance, during charged particle filtration, cf. \cite{Chowdiah1981}. Anderson \cite{Anderson1985} and Ajdari \cite{Ajdari1995} determined the vortex flow structure of an electrolyte in channels with nonuniform charged walls. The vortices are time-independent and have a fixed sense of circulation determined by the direction of the (steady) applied DC electric field. These studies were followed up by experiments on patterning flows in soft-lithography polydimethylsiloxane (PDMS) and glass wall channels \cite{Stroock2000,Stroock2003} establishing consistency of the experimental results with theoretical models.  More recently Kamsma \emph{et} al.  \cite{Kamsma2023} showed numerically, that an inhomogeneous wall charge distribution on a conical channel exhibited stronger current rectification and more pronounced memrisistive properties compared to uniformly charged conical channels \cite{Kamsma2023}, if the charge distribution changes sign when traversing the channel along its long axis (what they call ``bipolar'' surface charges). A theoretical framework for understanding the combined effect of active and charged boundaries relevant in biological or soft matter systems was recently developed in Ref. \cite{Shrestha2025b}.

The steady character of the aforementioned flows leads a microfluidic device to be classified as static or configurable such that the state 
of flow is either fixed at the manufacturing level and/or formatted to attain only limited predefined states  \cite{Paratore2022}. Current scientific and technological needs, however, require the state of devices to be \emph{reconfigurable} so to 
attain many different desired states ad voluntatem and in real time, cf. \cite{Shrestha2025}. To address this requirement we propose here a way to generate time-dependent vortex and other laminar flows by applying an alternating (AC) electric field on a rectangular channel or cylindrical capillary with modulated charged walls. As a result, this mode of reconfiguration generates a multitude of flow states that can be actuated as desired. This work suggests a reconfigurable way to tune and control electroosmotic flow in microfluidic devices that provides an efficient means for mixing, targeted solute transport, pumping and energy conversion. 

In this paper we show that, an AC electric field applied on electrolyte-filled channels or cylindrical capillaries with
insulating walls characterized by a modulated charge distribution, gives rise to oscillating vortex and other laminar flow patterns.  The vortex sense of circulation changes in time
with a period $2\pi/\omega$, where $\omega$ is the frequency of the applied AC field and this formulation lifts certain flow degeneracies associated with their time-independent counterparts. 
The flow is accompanied by an ionic current which exhibits hysteresis. 
The area of $I-V$ hysteresis loops is proportional to the energy dissipated during one cycle. 
This discussion is reminiscent of 
circuit elements that exhibit memory, that is, displaying properties depending on system state and history, as they were introduced in the seminal paper of Chua \& Kang \cite{Chua1976}. Following these early works we can associate with ``memory'' the inverse frequency giving the largest area in an $I(V )$ hysteresis loop. These observations may become essential in future unconventional concepts of memory \cite{Sangwan2020,Hadke2025}, and form the impetus for investigating control protocols of signal carriers.

This paper is organized as follows. In section \ref{sec: equations}
we analytically derive and display the ensuing streamline patterns for the geometries of a rectangular channel as well as a cylindrical capillary. We consider an electric field applied in a direction that is parallel to the wall charge variation. The time-dependence of the applied electric field gives rise to oscillating flows that lift the degeneracies associated with the time-independent case. This is analogous to the oscillating pressure gradient driven flow 
\cite[\S 24]{Landau1987}, but here the wall charge variation wavenumber and inverse Debye length are renormalized by the penetration depth.  The flow structure of cylindrical capillaries consists of oscillating tori. When the field is perpendicular to the wall charge variation, the streamline patterns obtained
resemble those determined in the experiments of Stroock \emph{et} al. \cite{Stroock2000}. The flows thus obtained here can be reconfigured as desired by changing the system parameters such as the frequency of oscillation.  
We demonstrate in section \ref{sec: unidirectional} that unidirectional electrolyte flow is possible
when the external DC bias is perpendicular to the wall charge variation. When they are parallel, the vortex flow vanishes after averaging over width of the channel. The advective current however is nonzero. This leads to the possibility of charge transfer in the absence of mass transfer. 
We finally derive the fluid flow structure for non-periodic wall charge distribution in Appendix \ref{sec: trigh}. 
In section \ref{sec: memory} we derive and display the longitudinal ionic advective current in the form of a current-voltage hysteresis loop. We thus determine a memory time from the frequency that maximizes the loop area. We interpret the system behavior with respect to a memory conductance that can become negative and diverge, in the spirit of similar works that have appeared in the literature \cite{Krems2010,Martinez2010,Wang2012,Luo2012}.

\section{\label{sec: equations}Electrolyte flow with modulated wall charge distribution}

It is well known \cite[\S 2.10]{Happel1965} that there are low Reynolds number
flows which are unsteady. This requires the ``vibrational'' Reynolds number $ \ell^2 \omega/\nu$
to not be small in comparison to the translational Reynolds number $\ell U/\nu$. Here $\ell, U$ and $\omega $ are characteristic 
length, translational velocity and frequency scales, respectively, while $\nu$ is the liquid kinematic viscosity. With
this understanding, the Navier-Stokes equations reduce to the form
\be \label{NS0}
\rho \partial_t \mathbf{v} = - \nabla p + \eta \nabla^2 \mathbf{v} + \mathbf{F} \quad \textrm{and} \quad 
\nabla  \cdot \mathbf{v} = 0, 
\ee
which can be employed to obtain estimates for the flow structure of time-dependent effects. Here, $p$ is
the hydrodynamic pressure, $\rho$ is the mass density of the liquid, $\eta$ is its dynamic viscosity and $\mathbf{F}$ a body force. In the context of this paper the body force is induced by an electric field acting on the free charges of an electrolyte.

We consider an external alternating electric field $\mathbf{E}$ with (real) frequency $\omega$ acting on a $1:1$ electrolyte in a rectangular channel and in a cylindrical capillary.
Because the walls are charged, an intrinsic electric field $-\nabla \phi$ forms (unrelated to the external field $\mathbf{E}$ in  this
discussion), providing the necessary equilibrium to the electrolyte charge species. When the potential $\phi$ is small (the Debye-H\"{u}ckel approximation), 
it satisfies the modified Helmholtz equation
\be\label{helm1}
\nabla^2 \phi = \kappa^2 \phi
\ee
where $\kappa^{-1}$ is the Debye length.
As in \citet{Ajdari1995}, it is implicit in \rr{helm1}
that the charge distribution in the electrolyte bulk is  given by $\rho_e = -\epsilon  \kappa^2 \phi$ where $\epsilon$ the dielectric constant
(a consequence of the Debye-H\"{u}ckel approximation, cf. Probstein \cite{Probstein1994}). 

The total electric field $\mathbf{E}-\nabla \phi $ exerts a body force $\mathbf{F} = \rho_e\left( \mathbf{E}-\nabla \phi\right) $ on the liquid electrolyte (fluid + ions), whose velocity $\mathbf{v}$ satisfies 
the Navier-Stokes equations \rr{NS0}
\be \label{NS1}
\rho \partial_t \mathbf{v} = - \nabla \tilde{p} + \eta \nabla^2 \mathbf{v} + \rho_e \mathbf{E} \quad \textrm{and} \quad 
\nabla  \cdot \mathbf{v} = 0, 
\ee
where $\tilde{p}$ is an effective
pressure. Since the expression $ \epsilon \kappa^2 \phi \nabla \phi$ is the gradient of a scalar function, this
contribution to the body force has been absorbed into the pressure and it does not affect the 
form of the streamlines and the commensurate velocity field. 

Motivation for studying this configuration stems from the experiments of Stroock \emph{et} al. \cite[figure 2]{Stroock2000}, who obtained
qualitatively similar flow patterns to the ones obtained here (albeit theirs being steady and expressed with respect to different boundary conditions). For clarity, our notation follows that of Ajdari \cite{Ajdari1995,Ajdari1996}
who showed that steady vortices, in a channel with $\cos q x$ wall charge variation, can form in a \emph{time-independent} electric field. In the adiabatic limit ($\omega \rightarrow 0$) our formulation recovers these time-independent results (this discussion is developed in Appendix \ref{sec: appendixajdari}). 
In addition, we adopt the notation of Landau \& Lifshitz \cite[p.89]{Landau1987} since our configuration features similarities 
with the corresponding problems of oscillating pressure-driven flows in a channel or capillary. For instance, the inertial term in \rr{NS0} leads
to the introduction of a penetration depth $\delta$ and complex wavenumber $k=(1+i)/\delta$, see Eq.~\rr{kdelta}. 
 
\subsection{\label{sec: par}Rectangular channel flow with applied field \emph{parallel} to surface charge variation}

\begin{figure}
	\vspace{5pt}
	\begin{center}
		\includegraphics[height=1.8in,width=5in]{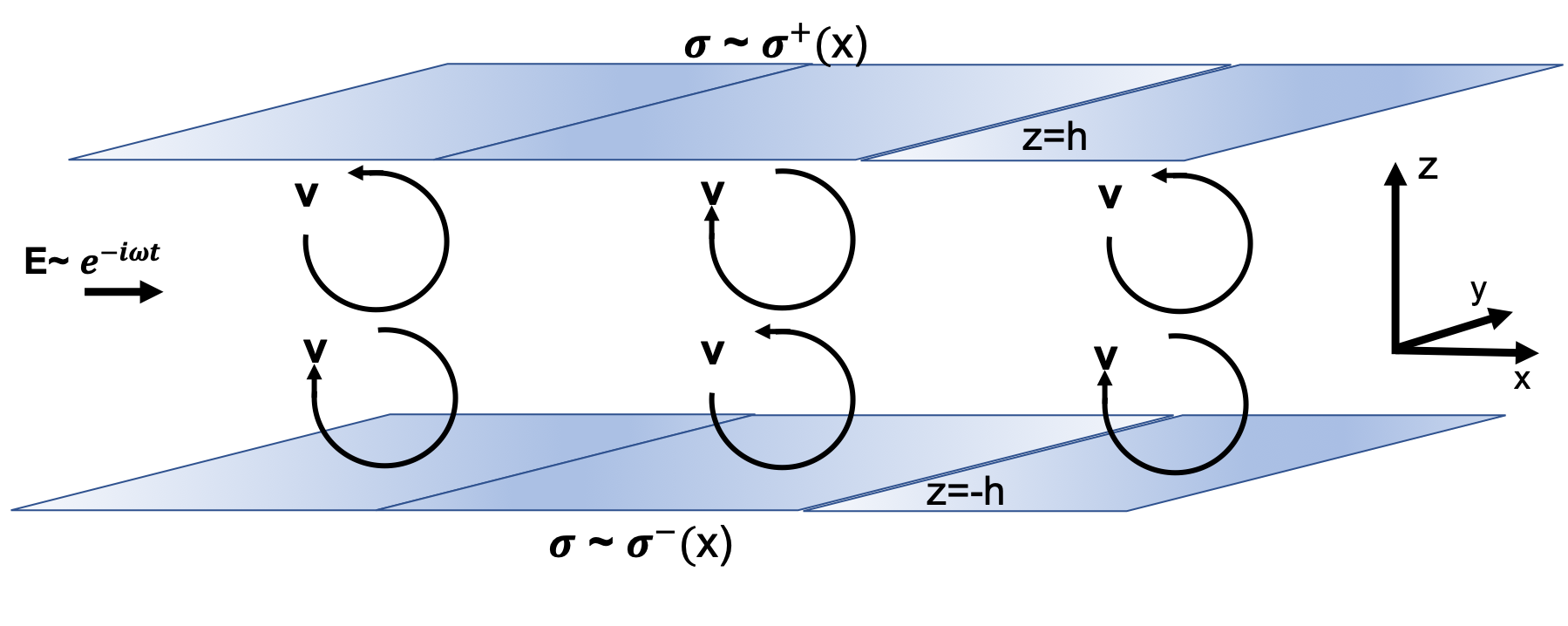}
		\vspace{-0pt}
	\end{center}
	\caption{AC electroosmosis with charge-modulated walls in a rectangular channel: A uniform alternating electric field $\mathbf{E} \sim e^{-i\omega t}\hat{\mathbf{x}}$ of frequency $\omega$ exerts a body force on a $1:1$ electrolyte in a rectangular channel of width $2h$
		with surface charge $\sigma^\pm = \sigma_0 \cos q x$ on the upper and lower insulating walls, respectively. 
		This gives rise to a system of vortices whose sense of rotation changes periodically in time according to the external electric field, cf. Fig. \ref{psi12}. 
		\label{channelAC1}  }
	\vspace{-0pt}
\end{figure}

We consider a channel of width $2h$ (see Fig. \ref{channelAC1})
whose  upper and lower walls are insulating and inhomogeneously charged with corresponding
distributions $\sigma^{\pm}(x)$ that only vary in the $x$-direction, satisfying the boundary conditions
\be\label{helmbc1}
\quad \partial_z \phi (x,z = \pm h) = \pm\frac{\sigma^{\pm}(x)}{\epsilon}
\ee
There is an externally applied field $\mathbf{E} = E_\parallel e^{-i\omega t}\hat{\mathbf{x}}$ parallel to the wall charge wave-vector $\mathbf{q}=q \hat{\mathbf{x}}$, and in the bulk there is the potential $\phi$ generated by the surface charges and ions.
The Navier-Stokes equations \rr{NS1} become
\be \label{NS2}
\rho \partial_t \mathbf{v} = - \nabla \tilde{p} + \eta \nabla^2 \mathbf{v} -\epsilon \kappa^2 \phi E_\parallel e^{-i\omega t}\hat{\mathbf{x}}. 
\ee
Following \cite{Ajdari1995} we consider two possible inhomogeneous surface charge distributions varying 
harmonically with wavevector $\mathbf{q} = q \hat{\mathbf{x}}$, giving rise to
the corresponding potentials as detailed below 
\be \label{phi1}
\sigma^\pm = \sigma_0 \cos q x, \quad \textrm{then} \quad \phi = \frac{\sigma_0} {\epsilon Q}\cos qx \frac{\cosh (Qz)}{\sinh(Qh)},
\ee
and 
\be \label{phi2}
\sigma^\pm = \pm \sigma_0 \cos q x, \quad \textrm{then} \quad \phi = \frac{\sigma_0} {\epsilon Q}\cos qx \frac{\sinh (Qz)} {\cosh(Qh)},
\ee
where the symbol $\sigma^\pm$ denotes surface charge distribution at the upper and lower channel walls, respectively,
(see Fig. \ref{channelAC1}), and 
\be \label{Q}
Q = (q^2  + \kappa^2)^{1/2}. 
\ee

To determine the velocity field we take the curl of the  Navier-Stokes equations \rr{NS2} and introduce the 
streamfunction $\psi$
\be \label{psi0}
\mathbf{v} = (u, 0, w) = (\frac{\partial \psi}{\partial z}, 0, -\frac{\partial \psi}{\partial x}).
\ee
The vorticity equation, written in terms of the streamfunction $\psi$, then takes the form
\be \label{vortpsi1}
\rho \partial_t \nabla^2 \psi = \eta \nabla^4 \psi - \epsilon \kappa^2 E_\parallel e^{-i\omega t} \partial_z \phi.
\ee
When the $x$-variation of the velocity field is proportional to $\cos q x$ and the time variation is harmonic with
frequency $\omega$ (instantaneously following the field), the streamfunction 
\be \label{psixzt}
\psi(x,z,t) = \psi(z) e^{-i\omega t}\cos qx 
\ee 
leads to the single equation for $\psi(z)$ { (with an abuse in notation)}
\be \label{Z4}
\partial_z^4 \psi+(K^2 - q^2) \partial_z^2 \psi -K^2 q^2 \psi  - \frac{\epsilon \kappa^2 E_\parallel}{\eta}  \partial_z \phi(z)=0,
\ee
where, with a slight abuse of notation, we've set  $\phi (x,z)=\phi(z) \cos qx$ in \rr{Z4}, and it is understood that $\phi(x,t)$ is given by one of \rr{phi1} or \rr{phi2}.
Here, 
\be \label{kdelta}
K = \sqrt{k^2  - q^2}, \quad 
i\omega = \nu k^2, \quad  k = \frac{1+i}{\delta}, \quad \delta = \sqrt{\frac{2\nu}{\omega}}, \quad \nu = \frac{\eta}{\rho}, 
\ee
is the familiar notation of the penetration depth $\delta$ and complex wavenumber $k$ of the 
oscillating pressure-driven flow in a channel, cf. \cite[p.84-89]{Landau1987} and $\nu$ is the kinematic viscosity of the 
electrolyte. Thus, the $z$ variation of $\psi(z)\sim e^{isz}$ has 
wavenumbers of the form
\be \label{s0}
s  = \pm i q, \quad \textrm{and} \quad s = \sqrt{ k^2 - q^2}\equiv K, 
\ee
leading to a solution of the form $
\psi(z) = A \cosh \! \left(q z \right)+B \sinh \! \left(q z \right)+C \cos \! \left(K z \right)+D \sin \! \left(K z \right) +\psi_p(z)
$, where $\psi_p$ is the particular solution associated with the last term in \rr{Z4} and $A, B, C$ and $D$ are integration constants to be determined by the no-slip boundary condition at the channel walls
{
\be
\psi(\pm h)= \frac{d \psi}{dz} (\pm h) =0. 
\ee
}
 For their calculation see Appendix \ref{sec: appendixajdari}.
Two features distinguish the eigenvalue set in \rr{s0} from previous studies. First, the oscillatory
character of our system has lifted the degeneracy
present in the time-independent counterpart (where  the four eigenvalues $s$ in \rr{s0} are instead two pairs of imaginary wavenumbers $\pm iq$, cf. \cite{Ajdari1995}). 
Second, the frequency of 
oscillation has introduced an additional length-scale to the problem (the penetration depth $\delta$, cf. Eq. \rr{kdelta})
which competes with the existing scales $h, \kappa^{-1}$ and $q^{-1}$. 

\begin{table}
	\begin{center}
		\def~{\hphantom{0}}
		\begin{tabular}{ll}
			\textrm{Quantity (units)}&			
			\textrm{Definition}\\
			$\omega\; (\textrm{s}^{-1})$ &  frequency of external field\\
			$\nu$ ($\textrm{m}^{2}\textrm{s}^{-1} $)    &       electrolyte kinematic viscosity \\
			$\delta = \sqrt{\frac{2\nu}{\omega}} \; (\textrm{m})$ &  liquid penetration depth (cf. Eq. \rr{kdelta}) \\
			$q \; (\textrm{m}^{-1})$ &  wall charge variation wavenumber\\
			$\kappa \; (\textrm{m}^{-1})$ &  inverse Debye length \\
			$Q=\sqrt{q^2 + \kappa^2} \; (\textrm{m}^{-1})$ & \\
			$k=\frac{1+i}{\delta} \; (\textrm{m}^{-1})$ &  liquid oscillating complex wave number (cf. Eq. \rr{kdelta}) \\
			$K = \sqrt{k^2 - q^2} \; (\textrm{m}^{-1})$ &  liquid complex wave number with charge variation (cf.  Eq. \rr{kdelta}) \\
			$\eta$ ($\textrm{kg}\:\textrm{m}^{-1}\textrm{s}^{-1} $)    &  electrolyte dynamic viscosity \\
			$2h, 2a$ (m) &  channel width, capillary diameter\\
			$\sigma^\pm \sim \sigma_0$  &  wall charge distribution\\
			$\epsilon $  &  dielectric constant\\
			$\phi $  &  electric potential \\
			$\psi $  &  streamfunction, cf. \rr{psi0}\\
			$p, \tilde{p}$  &  hydrodynamic and modified pressure\\
			$u,w$ (m/s) & channel horizontal and vertical electrolyte velocity components\\
			$v_r,v_z$ (m/s) &  capillary radial and axial electrolyte velocity components
		\end{tabular}
		\caption{\label{tab: table1}%
			Definitions of generic parameters and wavenumbers. $q, Q$ and $\kappa$ as in  \cite{Ajdari1995}. 
			$k$ and $\delta$ as in \cite[\S 24]{Landau1987}. 
		}
	\end{center}
\end{table}

\begin{figure}
	\vspace{5pt}
	\begin{center}
		\includegraphics[height=1.8in,width=5.8in]{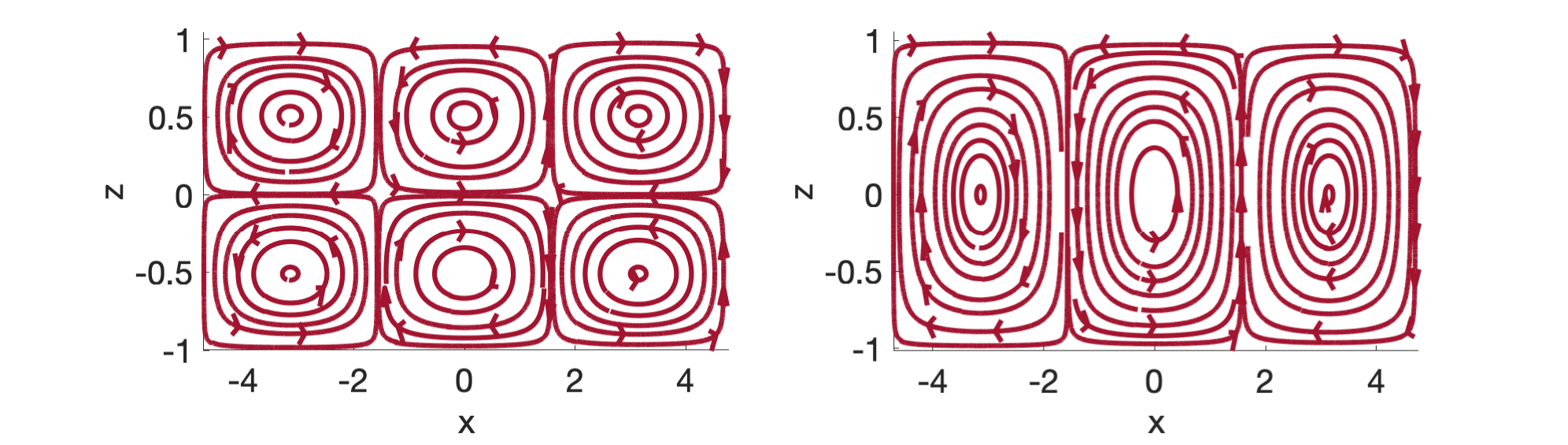}
		\vspace{-0pt}
	\end{center}
	\caption{Oscillating vortex structure in a channel of width $h=1$, induced by a uniform alternating electric field
		$\mathbf{E} = E_\parallel e^{-i\omega t}\hat{\mathbf{x}}$
		of frequency $\omega$ that is parallel to the wall charge variation wavevector $q \hat{\mathbf{x}}$,  cf. Fig. \ref{channelAC1} for the coordinate system. Left: dimensionless instantaneous streamlines  from  \rr{psiphi1} determined 
		for symmetric surface charges $\sigma^\pm = \sigma_0 \cos q x$ along the channel walls. Right: dimensionless instantaneous streamlines  from  \rr{psiphi2}, determined for 
		antisymmetric surface charges  $\sigma^\pm = \pm \sigma_0 \cos q x$. Compare with the upper and lower row
		of \cite[figure 2]{Ajdari1995}. It is thus clear that  the pattern displayed on the left smoothly deforms and becomes the one
		to the right by continuously varying the relative phase of the charge distribution between the upper and lower channel 
		walls (see discussion in the main text). 
		\label{psi12}  }
	\vspace{-0pt}
\end{figure}

For instance, in the symmetric charge distribution case leading to the electric potential \rr{phi1}, the streamfunction \rr{psixzt} satisfying the vorticity 
equation \rr{vortpsi1}, prompts the function $\psi(z)$ (by solving Eq. \rr{Z4}) to acquire the form
\be \label{psiphi1}
\psi(z) = \frac{\sigma_0E_\parallel \sinh \! \left(Q z \right) }{\eta (\kappa^2+k^2) \sinh(Qh)} - A \sinh \! \left(q z \right)- B \sin \! \left(K z \right) 
\ee
with 
\be \label{ABchannel}
\left( \begin{array}{c}
	A \\
	B
\end{array} \right) = 
\frac{\sigma_0E_\parallel}{\eta (\kappa^2+k^2) \sinh(Qh)}
\left( \begin{array}{c}
	\frac{K\cos \! \left(K h \right) \sinh \! \left(Q h \right)  - Q\sin \! \left(K h \right) \cosh \! \left(Q h \right)  }{K\sinh \! \left(q h \right)  \cos \! \left(K h \right) - q \cosh \! \left(q h \right)  \sin \! \left(K h \right)} \\
	\frac{Q\cosh \! \left(Q h \right) \sinh \! \left(q h \right)  - q\sinh \! \left(Q h \right) \cosh \! \left(q h \right)  }{K\sinh \! \left(q h \right)  \cos \! \left(K h \right) - q \cosh \! \left(q h \right)  \sin \! \left(K h \right)}
\end{array} \right).
\ee
This corresponds to the velocity components 
\be \label{vphi1}
u(x,z,t) = \left[  \frac{\sigma_0E_\parallel Q\cosh \! \left(Q z \right) }{\eta (\kappa^2+k^2) \sinh(Qh)}-A q \cosh \! \left(q z \right)- B K \cos \! \left(K z \right)  \right] e^{-i\omega t}\cos qx
\ee
and
\be \label{wphi1}
w(x,z,t) = \left[ \frac{\sigma_0E_\parallel \sinh \! \left(Q z \right) }{\eta (\kappa^2+k^2) \sinh(Qh)}- A \sinh \! \left(q z \right)- B \sin \! \left(K z \right)   \right] e^{-i\omega t}q \sin qx.
\ee
In Fig. \ref{psi12} we display the dimensionless instantaneous streamlines  from  \rr{psiphi1} (left) and \rr{psiphi2} (right) associated, respectively, with
the symmetric surface charges $\sigma^\pm = \sigma_0 \cos q x$ and
antisymmetric surface charges  $\sigma^\pm = \pm \sigma_0 \cos q x$, along  the channel walls. 
It is easy to see that the transition from the left pattern to the right occurs as the relative phase of the charge distribution on
each wall is continuously varied from (in the notation of  \cite{Ajdari1995}) $\gamma^- - \gamma^+ =0$ to $\pi$. 
See the third column in  \cite[figure 2]{Ajdari1995}.

We introduce the velocity scaling factor
\be \label{scaled}
u_0 = \frac{\sigma_0 E_\parallel }{\eta \kappa}. 
\ee
The unsteady streamfunction \rr{vphi1} reduces to its static counterpart \rr{ajdaripsi} in the limit $\omega \rightarrow 0$. A distinguishing feature of the solution \rr{psiphi1} in comparison with previous studies cf. Appendix \ref{sec: appendixajdari} is 
that it has lifted another degeneracy; the one that exists in the time-independent problem when 
$\kappa \rightarrow 0$ (although this limit is not expected to practically ever be reached), see the front factors of Eq. \rr{ajdaripsi}-\rr{ajdariD} in comparison to the front factors of \rr{psiphi1} with \rr{ABchannel}. 

\subsection{\label{sec: cylinder}Cylindrical capillary flow with axially applied field}
\begin{figure}
	\vspace{5pt}
	\begin{center}
		\includegraphics[height=1.6in,width=4in]{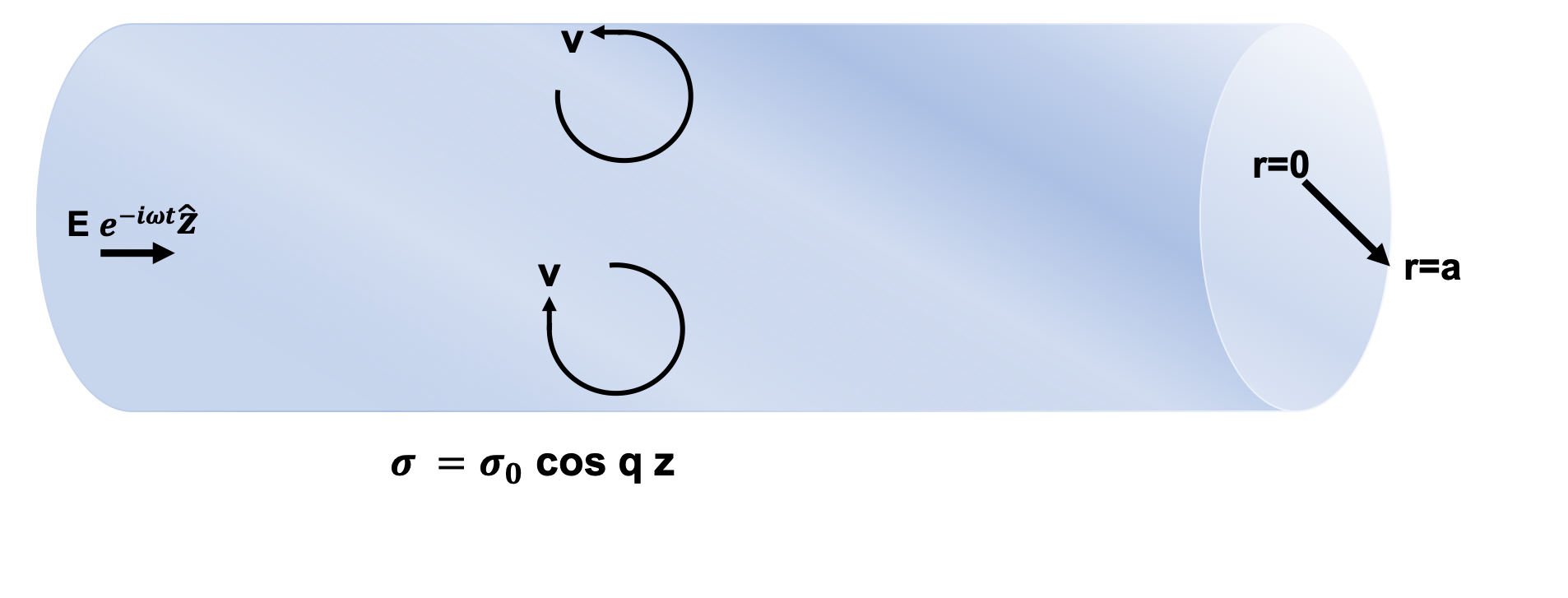}
		\vspace{-0pt}
	\end{center}
	\caption{AC electroosmosis in a cylindrical capillary with charge-modulated walls: A uniform alternating electric field $\mathbf{E} = Ee^{-i\omega t}\hat{\mathbf{z}}$ of frequency $\omega$ exerts a body force on a $1:1$ electrolyte in a cylindrical capillary of radius $a$
		with inhomogeneous surface charges, $\sigma = \sigma_0 \cos q z$ on its insulating wall. 
		This gives rise to a system of oscillating tori (whose cross-section are the displayed vortices) whose sense of rotation changes according to the external field's, cf. Fig. \ref{psi12cyl} for a calculated cross-section. 
		Compare with figures \ref{channelAC1} and \ref{psi12} of the channel case. 
		\label{cylinder1ac}  }
	\vspace{-0pt}
\end{figure}

The above analysis can be repeated for the case of a cylindrical 
capillary with circular cross-section of radius $a$ whose insulating wall carries a charge
\be
\sigma = \sigma_0 \cos qz, \quad \textrm{at} \quad r=a, 
\ee
cf. Fig. \ref{cylinder1ac}. 
Again we consider the Debye-H\"{u}ckel approximation where $\rho_e = - \epsilon \kappa^2 \phi$. 
Assuming  $\phi(r,z) = \phi(r) \cos qz $, with a slight abuse of notation, Gauss law reduces to
\be \label{rhophicylinder}
\partial_r^2 \phi + \frac{1}{r}\partial_r \phi  - (\kappa^2 +q^2) \phi = 0, \quad \textrm{subject to} \quad \partial_r \phi (a) = \frac{\sigma_0}{\epsilon}.
\ee

Thus, the potential acquires the form
\be \label{phirz}
\phi(r,z) = \frac{\sigma_0}{\epsilon} \frac{I_0(Qr)}{QI_1(Qa)}\cos qz
\ee
where $I_0$ and $I_1$ are modified Bessel functions of the first kind and 
with $Q^2 = \kappa^2 +q^2$, defined in \rr{Q}.

If a uniform alternating electric field is applied along the axis of the cylinder $\mathbf{E} = E e^{-i\omega t}\hat{\mathbf{z}}$
(cf. Fig. \ref{cylinder1ac}), it exerts a torque on the liquid of the form
\be \label{lintorque}
\nabla \times (\rho_e \mathbf{E}) = \epsilon \kappa^2 \partial_r \phi  E \hat{\bm{\phi}}, 
\ee
where $\hat{\bm{\phi}}$ is the azimuthal unit vector in a system of circular cylindrical coordinates. 
The vorticity equation in terms of a streamfunction $\psi(r,z,t)$, expressed in cylindrical coordinates is well known cf. 
\cite[\S 4.7]{Happel1965} and we briefly revisit the salient points of its derivation. 
The vorticity equation can be written in the form
\be \label{vortcyl1}
\rho \partial_t \nabla \times \mathbf{v} = -\eta \nabla \times \nabla \times  \nabla \times \mathbf{v} + \epsilon \kappa^2 \partial_r \phi  E \hat{\bm{\phi}}.
\ee
The flow is axisymmetric and only the radial and axial velocity components $v_r$ and $v_z$, respectively, are non-vanishing.
With a Stokes streamfunction $\psi(r,z,t)$ satisfying
\be \label{Stokespsi}
v_r = \frac{1}{r} \frac{\partial \psi}{\partial z}, \quad v_z =- \frac{1}{r} \frac{\partial \psi}{\partial r}, 
\ee
the incompressibility condition is automatically satisfied and 
the $\hat{\bm{\phi}} $ component of the vorticity can be written in the form 
\be
r (\nabla \times \mathbf{v}) \cdot \hat{\bm{\phi}}  = \left( \partial_r^2 - \frac{1}{r} \partial_r + \partial_z^2  \right) \psi \equiv \mathcal{D}^2 \psi
\ee
and this defines the operator $\mathcal{D}^2$ (denoted by the symbol $E^2$ in \cite{Happel1965}, which however here is reserved to denote the electric field amplitude). It is then easy to show that the vorticity equation \rr{vortcyl1} becomes, cf. 
\cite[Eq. (4-7.12)]{Happel1965} 
\be
\frac{1}{\nu}\partial_t \mathcal{D}^2 \psi = \mathcal{D}^4 \psi + \frac{\sigma_0\kappa^2 E}{\eta}\frac{rI_1(Qr)}{I_1(Qa)} e^{-i\omega t} \cos qz.
\ee
Assuming a streamfunction of the form $\psi(r,z,t) = \psi(r) e^{-i\omega t}\cos qz$ (with a slight abuse of notation), its radially dependent factor satisfies 
{
\be \label{psi4r}
\left[\frac{d^{2}}{d r^{2}}-\frac{1 }{r}\frac{d}{d r} - q^{2} \right]\left[ \frac{d^{2}}{d r^{2}}- \frac{1}{r}\frac{d}{d r} + K^2\right]\!\psi + \frac{\sigma_0\kappa^2 E}{\eta}\frac{rI_1(Qr)}{I_1(Qa)}
=0,
\ee
}
which is the counterpart of \rr{Z4}, $Q^2 = \kappa^2 +q^2$ and $K^2 = k^2 -q^2$ defined in \rr{Q} and \rr{kdelta}, respectively.  
The general solution of \rr{psi4r}, that is finite at $r=0$, is of the form
\be \label{psir}
\psi(r) = r \left[A I_1(qr) + B J_1(Kr)  -\frac{\sigma_0 {E} I_{1}\! \left(Q r \right) }{\eta  I_{1}\! \left(Q a \right)  \left(k^{2}+\kappa^{2}\right)} \right],
\ee
where $J_1$ is the Bessel function of the first kind and $A$ and $B$ are arbitrary constants to be determined by the boundary conditions. These are the no-slip conditions at the 
capillary walls
\be
\psi(a)= \frac{d \psi}{dr} (a) =0. 
\ee
The two constants are 
\be \label{AB}
\left( \begin{array}{c}
	A \\
	B
\end{array} \right) = 
\frac{\sigma_0E}{\eta (\kappa^2+k^2) I_1(Qa)}
\left( \begin{array}{c}
	\frac{KJ_0(Ka)I_1(Qa) - QJ_1(Ka)I_0(Qa)}{KJ_0(Ka)I_1(qa) - qI_0(qa)J_1(Ka)} \\
	\frac{QI_1(qa)I_0(Qa) -q I_1(Qa)I_0(qa)}{KJ_0(Ka)I_1(qa) - qI_0(qa)J_1(Ka)}
\end{array} \right), 
\ee
where $J_0$ is the Bessel function of the first kind, of zeroth order.

\begin{figure}
	\vspace{5pt}
	\begin{center}
		\includegraphics[height=2in,width=4in]{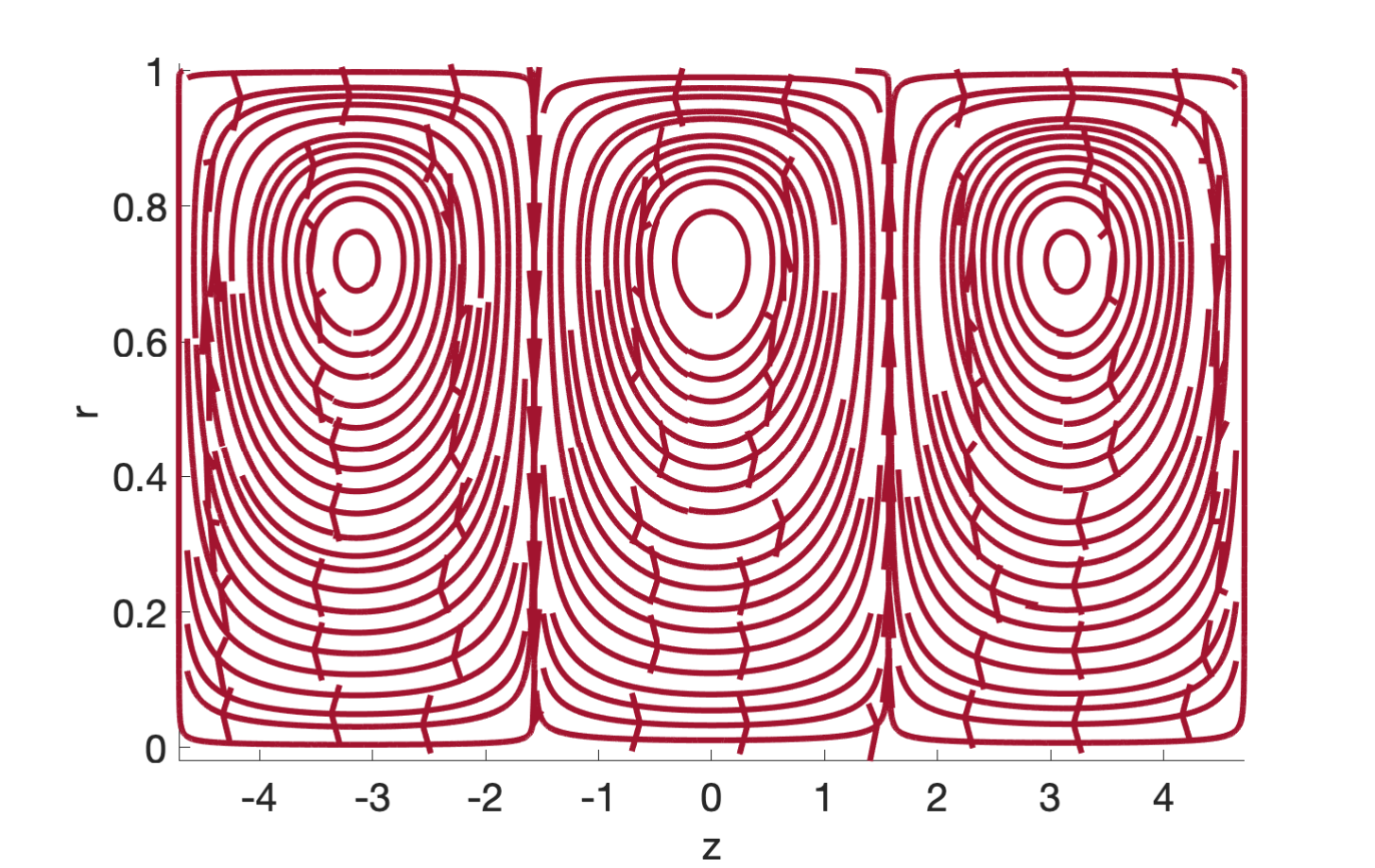}
		\vspace{-0pt}
	\end{center}
	\caption{Oscillating toroidal vortex structure in a capillary of radius $r=1$, induced by a uniform alternating electric field
		$\mathbf{E} = Ee^{-i\omega t}\hat{\mathbf{z}}$
		of frequency $\omega$ that is parallel to the wall charge variation wavevector $q \hat{\mathbf{z}}$ (parallel to the cylinder axis),  cf. Fig. \ref{cylinder1ac} for the coordinate system. Here  we display an $r-z$ slice of the dimensionless instantaneous streamlines $\psi(r,z,t) = \psi(r) e^{-i\omega t}\cos qz$ with $\psi(r)$ from  \rr{psi4r} and \rr{AB} determined 
		for charges $\sigma= \sigma_0 \cos q z$ on the capillary wall located at $r=1$. 
		\label{psi12cyl}  }
	\vspace{-0pt}
\end{figure}

In Fig. \ref{psi12cyl} we display a cross-section of the flow in the $r-z$ plane. The flow consists of identical vortices, which, when extended along the azimuthal direction $\hat{\bm{\phi}} $, form tori which alternate in their sense of circulation as one moves parallel to the axis of the cylinder (in the $z$-direction). 
The circulation direction changes periodically depending on the orientation of the electric field $\mathbf{E} = E e^{-i\omega t}\hat{\mathbf{z}}$. Employing \rr{Stokespsi} and \rr{AB} shows 
that the radial velocity $v_r$ is proportional to $q$ and thus vanishes when the wall charges are spatially uniform.

\begin{figure}
	\vspace{5pt}
	\begin{center}
		\includegraphics[height=3.3in,width=6.8in]{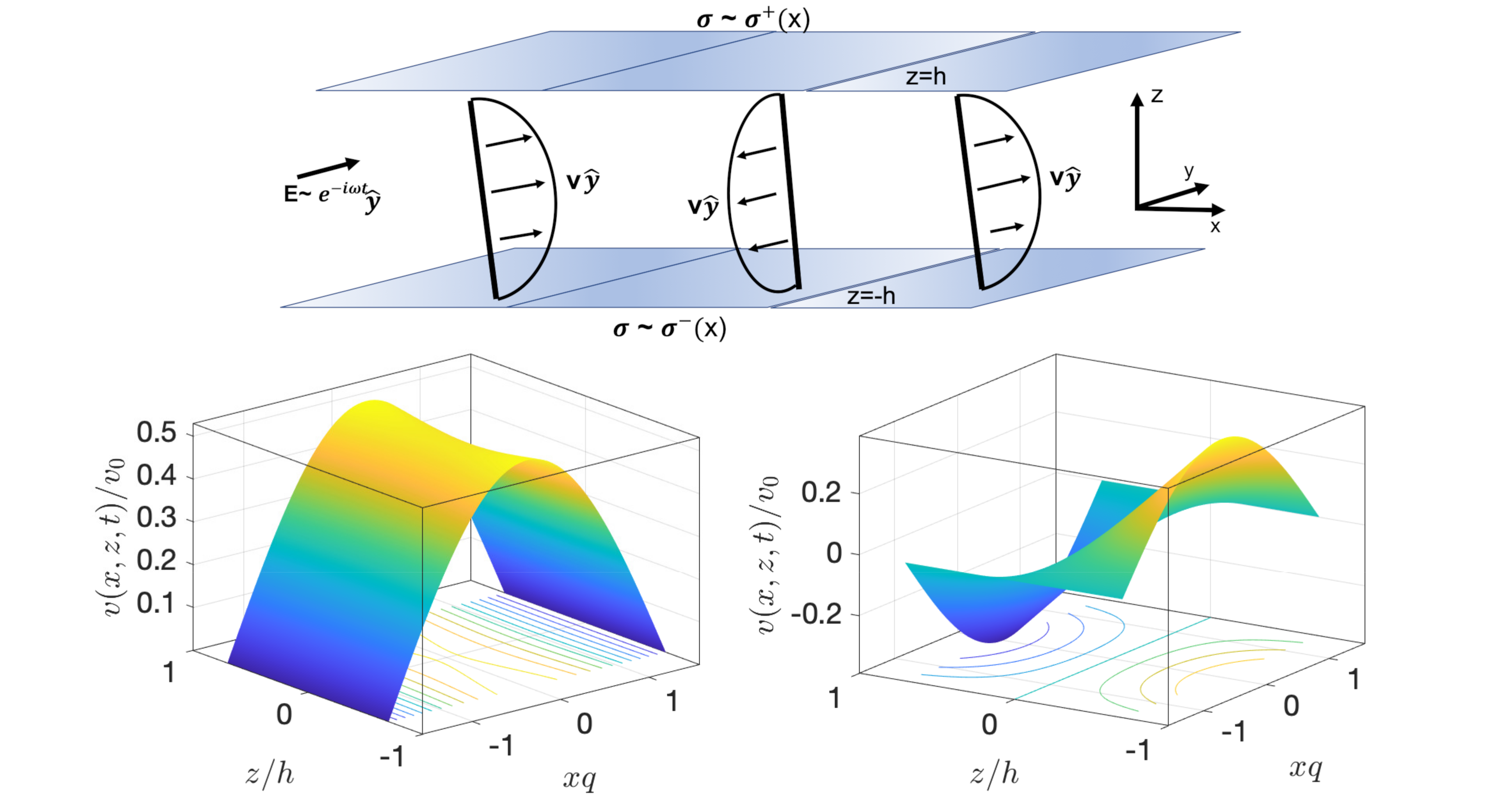}
		\vspace{-0pt}
	\end{center}
	\caption{AC electroosmosis in a rectangular channel with applied field \emph{perpendicular} to wall charge variation. Top: An alternating electric field $\mathbf{E} \sim e^{-i\omega t}\hat{\mathbf{y}}$ of frequency $\omega$ exerts a body force on a $1:1$ electrolyte in a rectangular channel of width $2h$
		with symmetric inhomogeneous wall charges, $\sigma^\pm = \sigma_0 \cos q x$ 
		on the upper and lower insulating walls. 
		Over each tile, the charge distribution gives rise to a non-vanishing velocity averaged over the channel width \rr{v1}. Bottom: Oscillating velocity profile in a channel of width $2h$, induced by a uniform alternating electric field
		$\mathbf{E} = E_\perp e^{-i\omega t}\hat{\mathbf{y}}$
		of frequency $\omega$ that is \emph{perpendicular} to the wall charge variation for the coordinate system. Snapshot of the velocity profile $v(x,z,t)\hat{\mathbf{y}}$ from  \rr{v1} at time $t=1$
		when the surface charges at the chanel walls $z=\pm h$ are symmetric
		$\sigma^\pm = \sigma_0 \cos q x$ (left). Snapshot of the velocity profile $v(x,z,t)\hat{\mathbf{y}}$ (by exchanging the cosines with sines in \rr{v1}, and vice-versa) at time $t=1$
		when the surface charges at the chanel walls $z=\pm h$ are antisymmetric
		$\sigma^\pm =\pm  \sigma_0 \cos q x$ (right). 
		\label{vphi1phi2}  }
	\vspace{-0pt}
\end{figure}

\subsection{\label{sec: perp}Rectangular channel flow with applied field \emph{perpendicular} to surface charge variation}

Next consider an external alternating electric field that is applied perpendicular to wall charge variation $\mathbf{E} = E_\perp e^{-i\omega t}\hat{\mathbf{y}}$ with symmetric or antisymmetric wall distributions given corresponding potentials Eq.~\rr{phi1} and \rr{phi2}, satisfying Eq. \rr{helm1} in the Debye-H\"{u}ckel approximation as in the previous section. This configuration represents the oscillatory counterpart of Ajdari \cite[Eq. (9) \& (10)]{Ajdari1995}. 

The momentum equation becomes in this case
\be \label{NS3}
\rho \partial_t \mathbf{v} = - \nabla \tilde{p} + \eta \nabla^2 \mathbf{v} -\epsilon \kappa^2 \phi E_\perp e^{-i\omega t}\hat{\mathbf{y}},
\ee
{where $\mathbf{v}(x,z) = (u(x,z),v(x,z),w(x,z))$ in Cartesian coordinates. 
The vorticity equation shows that 
\be \label{NS4}
\left( \partial_x, \partial_z\right)  \left[\rho \partial_t v - \eta \nabla^2 {v} -\epsilon \kappa^2 \phi E_\perp e^{-i\omega t}\right]=(0,0),
\ee
and that $\left( \rho \partial_t  - \eta \nabla^2\right) (\partial_z u - \partial_x w) =0$. As in \cite[Eq. (9) \& (10)]{Ajdari1995},  a solution that satisfies no-slip boundary conditions at $z =\pm h$ is $u=w \equiv 0$.  The velocity component $v$ satisfying $\rho \partial_t v - \eta \nabla^2 {v} -\epsilon \kappa^2 \phi E_\perp e^{-i\omega t }=0$
\rr{NS4} obtains the form $v(x,z, t) = Z(z)  e^{-i\omega t}\cos qx$, thereby }
\be \label{Zcosh}
Z'' + (k^2  - q^2) Z - \frac{\sigma_0 \kappa^2 E_\perp}{\eta Q}  \frac{\cosh (Qz)}{\sinh(Qh)} =0
\ee
where we employed the electric potential \rr{phi1} determined when both walls
carry the same charge ($\sigma_0 \cos qx$). We recover the case where the electric potential employed is \rr{phi2} (determined when the walls
carry opposite charge), by performing the exchange $\cosh \leftrightarrow \sinh$ in \rr{Zcosh}.  

The resulting velocity is 
\be \label{v1}
v(x,z,t) = \frac{\kappa^2 E_\perp \sigma_0\cos qx}{\eta (k^2 + \kappa^2) Q\sinh (Qh)} \left( \cosh (Qz) - \frac{\cosh (Qh)}{\cos (Kh)} \cos (Kz )\right)e^{-i\omega t}
\ee
where $K$ was defined in \rr{kdelta}. The velocity profile that corresponds to the electric potential in Eq. \rr{phi2}, is obtained from \rr{v1} by performing the exchange
$\cosh \leftrightarrow \sinh$ and $\cos \leftrightarrow \sin$, leaving $\cos qx$ intact. It is clear that \rr{v1} recovers previous static results in the adiabatic limit $\omega \rightarrow 0$. Thus, setting
$K = \sqrt{-q^2}$ in \rr{v1} and employing the identity $\cos Kz/\cos Kh \equiv \cosh qz /\cosh qh$, we recover 
the (static) velocity profile of Ajdari \cite[Eq. (26)]{Ajdari1996}. 

A snapshot of the velocity profile \rr{v1}, scaled by the factor \rr{scaled} which here is denoted by $v_0$, that is,
\be \label{v0}
v_0 = \frac{\sigma_0 E_\perp}{\eta \kappa},
\ee
is displayed in the bottom left panel of Fig. \ref{vphi1phi2} over half period ($\pi/q$) in the $x$-direction. This is reminiscent of the experimental results of Stroock \emph{et} al.  \cite[figure 2]{Stroock2000}. Moreover,  it is seen that the profile is oscillating but of single sign (in the chosen interval of the $x$-axis) and can thus
be employed for unidirectional transport of electrolytes, as discussed in the following section.  The bottom right panel of Fig. \ref{vphi1phi2} displays the scaled velocity profile by $v_0$, when the upper and lower channel walls carry 
charges of opposite sign. Here the flow, averaged over the width of the channel, vanishes.  Such a pattern can also be developed for a symmetric
$\sigma^\pm = \sigma_0 \cosh q x$ wall charge distribution (cf. Fig. \ref{vcoshsinh} in Appendix \ref{sec: trigh}).

\section{\label{sec: unidirectional}AC-induced \emph{unidirectional} electrolyte motion in channels and capillaries}

In addition to the vortex flow patterns of the previous sections, it is of practical interest to examine the presence of  unidirectional electrolyte transport. In this section we calculate the velocity averaged over the channel width for the perpendicular field to the charge variation, and examine their limiting behaviors. The averaged velocity is
\be
\langle  v  \rangle = \frac{1}{2h} \int_{-h}^h v dz.
\ee
Consider the case of uniform wall charges $\sigma^\pm = \sigma_0$ along both walls. The motivation for the choice of 
symmetric wall charge comes from the observation that an optimal unidirectional velocity 
is expected to arise when the wall excitations are identical and of the same phase \citep{Yeh2011}. The configuration where the velocity does not vanish when averaged over the channel width, is the one with symmetric charges along the channel walls.

\subsection{\label{sec: unidirectionalchannelperp} Average velocity in rectangular channel with field perpendicular to charge variation} 
In the
case of the perpendicular applied field, the velocity profile \rr{v1} associated with the symmetric charge on the walls $z= \pm h$, gives rise to unidirectional charge transport. Averaging \rr{v1} over the channel width we obtain
\be \label{v1av}
\langle v \rangle \sim\frac{\sigma_0 E_\perp e^{-i\omega t} }{\eta h (k^2 + \kappa^2)}  \begin{cases}  \frac{ \kappa^2 }{Q^2}  \left[ 1- \frac{Q}{K} \tan (Kh)\coth (Qh)
\right] \cos qx \,, 
& \quad \textrm{for} \quad q \neq 0 \\
1 - \frac{\kappa}{k} \tan(kh) \coth(\kappa h)  \,,
& \quad \textrm{for} \quad q \rightarrow 0.
\end{cases}
\ee
In the left two panels of figure \ref{newplot} we display the absolute value of the amplitude contained in the square brackets of \rr{v1av}, for various values of the wall charge variation wavenumber $q$. 

Of interest is also the adiabatic limit $\omega \rightarrow 0$ (and thus $k\rightarrow 0$). 
Defining the Langevin function 
\be \label{L}
\mathscr{L}(x) = \coth  x - \frac{1}{x}, 
\ee
the $q\rightarrow 0$ limit of \rr{v1av} becomes 
\be \label{vomegalow}
\langle  v \rangle \sim - \frac{\sigma_0E_\perp}{\eta \kappa} \mathscr{L} (\kappa h)
\quad \textrm{at} \quad q =0 \quad \textrm{and} \quad \omega \rightarrow 0, 
\ee
(over times which satisfy $\omega t \ll 1$).
Since the argument $\kappa h$ in this paper is considered to be rather large, the Langevin function appearing in \rr{vomegalow} is approximately 
equal to $1$. The resulting velocity $\langle  v \rangle \sim - \frac{\sigma_0E_\perp}{\eta \kappa}$ is of a type obtained in 
standard (static) electroosmosis formulations, cf. \cite[p. 10.24]{Melcher1981}. We note that this is a plug-like induced by the  large values of the inverse Debye wavelength $\kappa$ and it differs from the plug-like flow that forms when $\omega \rightarrow \infty$, cf. Appendix \ref{sec: appendixPressure}. 

In the right panel of figure \ref{newplot} we display the absolute value of the amplitude contained in \rr{v1av} for $q\rightarrow 0$
versus the scaled frequency $\frac{\omega}{\nu \kappa^2}$ where the velocity scale $v_0$ was defined in \rr{v0}. 
For low scaled frequency values the velocity amplitude reaches a plateau (to the left of each curve) denoting the plug-like adiabatic limit \rr{vomegalow}, expressed with respect to a Langevin function \rr{L} and resembles the standard (static) electroosmosis expression met in the literature.  
For large frequencies, in principle, the rapid oscillation of the complex exponential $e^{-i\omega t}$ should also be taken into account.

\subsection{\label{sec: estimates}Estimates}
\begin{figure}
	\vspace{5pt}
	\begin{center}
		\includegraphics[height=2.4in,width=7in]{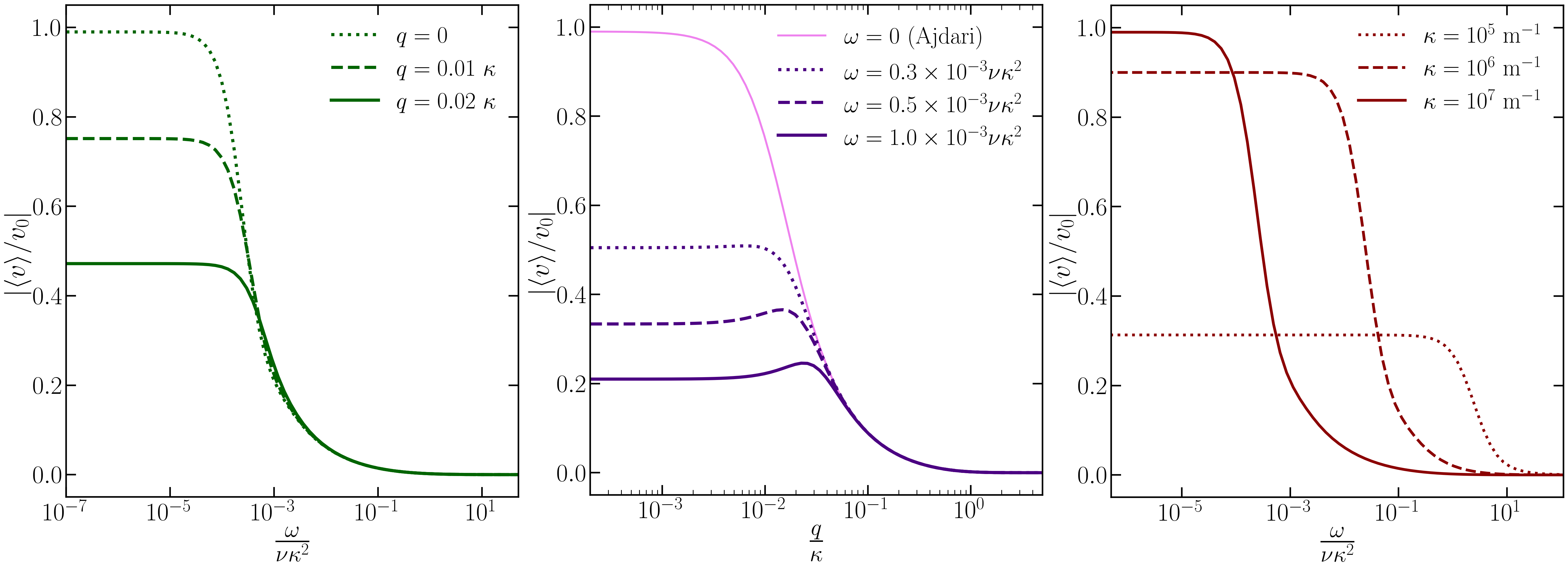}
		\vspace{-0pt}
	\end{center}
	\caption{{Unidirectional channel averaged velocity with perpendicular applied AC field. Left panel displays the velocity amplitude versus the scaled frequency $\frac{\omega}{\nu \kappa^2}$ for various wavenumbers $q$ of wall charge variation, 
		where $
		v_0 = \frac{\sigma_0 E_\perp}{\eta \kappa} 
		$. The plateau at the low frequency values denotes the plug-like adiabatic limit. Middle panel: Amplitude of the scaled unidirectional velocity \rr{v1av} for the perpendicular case as a function of $q/\kappa$ for different values of frequency $\omega$, recovering the static case of Ajdari \cite{Ajdari1996} in the adiabatic limit. The right panel displays the velocity \rr{v1av} in the $q\rightarrow 0$ limit. } 
		\label{newplot}  }
	\vspace{-0pt}
\end{figure}

From figure \ref{newplot} and the long wavelength expression
\rr{vomegalow}  it can be seen that for low frequencies the electrolyte velocity is of the order
of the scaled parameter $v_0$ defined in \rr{v0}. For higher frequencies the velocity magnitude is reduced. Thus, we can employ
\be
v\sim v_0 = \frac{\sigma_0E}{\eta \kappa},
\ee
 as an upper bound that the electrolyte velocity can attain. 
Employing standard parameter values for water $\epsilon = 7\times 10^{-10}$ F/m, $\eta = 0.001 \textrm{kg/(m}\cdot \textrm{s)}$, 
$\kappa = 10^{7}\textrm{ m}^{-1}$ and $E =  9.5\times 10^4$ V/m (as in \cite{Stroock2000}) and a surface charge with the nominal value
$\sigma_0 = \epsilon E$
we obtain the estimate
\be
v\sim 0.63 \textrm{ mm/s}.  
\ee 
This value is of the same order of magnitude as standard electroosmosis velocity estimates 
(cf. Probstein \cite[below (6.5.5)]{Probstein1994}) and one order of magnitude higher than those
of the experiments of Stroock \emph{et} al. \cite{Stroock2000}.

\subsection{\label{sec: unidirectionalchannelpar} Unidirectional velocity in rectangular channel with field parallel to charge modulation} 
{
In the limit  $q\rightarrow 0$ the velocity in the rectangular and cylindrical cases becomes unidirectional and oscillatory in the $\hat{\mathbf{x}}$ direction (channel case) and in the $\hat{\mathbf{z}}$ direction (cylinder case). When averaged over the width of the channel/radius of cylinder, they both vanish. The convective current, however, is non-zero for all values of $q$.  This observation implies that the oscillatory electric field in a capillary with wall charge variation gives rise to hysteretic flow (evident from the presence of the renormalized Stokes wavenumber $K$ in \rr{kdelta}) and is thus endowed with traces of memory, a subject we develop in the ensuing sections. }

\section{\label{sec: memory}Memory Effects}
Recent work has focused on imitating physiological processes taking place in a synapse to create highly efficient iontronic computing devices, carrying information via ion transport in an aqueous environment rather than via electrons and holes in the crystal lattice, and will thus be manipulated by both chemical and electrical signals.

The plot of $I$ vs. $V$ in an iontronic device shows a hysteresis effect cf.  \citep[Fig. S13-S15]{Kamsma2024}, and this is interpreted as sign of ``memory''. The largest the enclosed area, the largest the memory is. The area of the hysteresis loops can be calculated as a function of the frequency $\omega$ of the applied voltage. The frequency that maximizes the area of the loop determines the characteristic `memory' time $\tau_M$, cf. \citep[Eq.(3) \& Fig. 3C]{Robin2023} and 
\citep[Fig. 3(c)]{Kamsma2023}.

For the static case, Osterle and coworkers \cite{Morrison1965} pointed out that although the current in the transverse direction to the capillary walls vanishes, the current in the longitudinal direction (parallel to the walls) is nonzero, cf. \cite[Eq. (23)]{Morrison1965}. This remains true in the present oscillatory configuration. We take advantage of this concept to show that rectangular channels or cylindrical capillaries in the presence of wall charge variation give rise to a current, which is hysteretic with respect to the applied voltage.

Hysteresis loops in the literature appear to be pinched (current vanishes when the voltage does). It was however argued by Pershin, Di Ventra and coworkers \cite{Krems2010,Martinez2010,Pershin2019} that pinching is not a necessary defining characteristic displayed by a system with memory. In the context of ionic systems, it is evident from the experimental figures of \citet[Fig. S13-S15]{Kamsma2024} that hysteresis loops can be open. Open hysteresis loops, displaying capacitive behavior, can also be seen in recent theoretical studies, for instance, in Figures 6Aa/d and 11B of \cite{Bisquert2024}. Below, we show that the natural behavior of an $I-V$ hysteresis loop emanating from the advective longitudinal current, within our approximate formulation, is not pinched.

Negative resistance or differential resistance has been documented in a number of recent works
\cite{Luo2012,Wang2012,Perez2021,Bisquert2023,Yang2024,Bisquert2024}. Likewise, we will here show that the conductance displayed by our ionic system can become negative and divergent as the external voltage approaches zero. This behavior is reminiscent of capacitive systems, for instance see Fig. 3 of \cite{Krems2010} and 
Figure 3B of \cite{Wang2012}, both showing negative capacitance in experiments whose behavior resembles the right-most panel of Fig.  \ref{memory_channel}.

For discussion in this memory section, below we will need the following current and voltage scalings
\be\label{I0V0}
I_0  = \frac{\sigma_0^2 E}{\eta}, \quad \textrm{and} \quad V_0 = EL,
\ee
where $L$ is the length of the channel or cylindrical tube over which the voltage is applied.

\begin{figure*}
\vspace{5pt}
\begin{center}
\includegraphics[height=3in,width=4.8in]{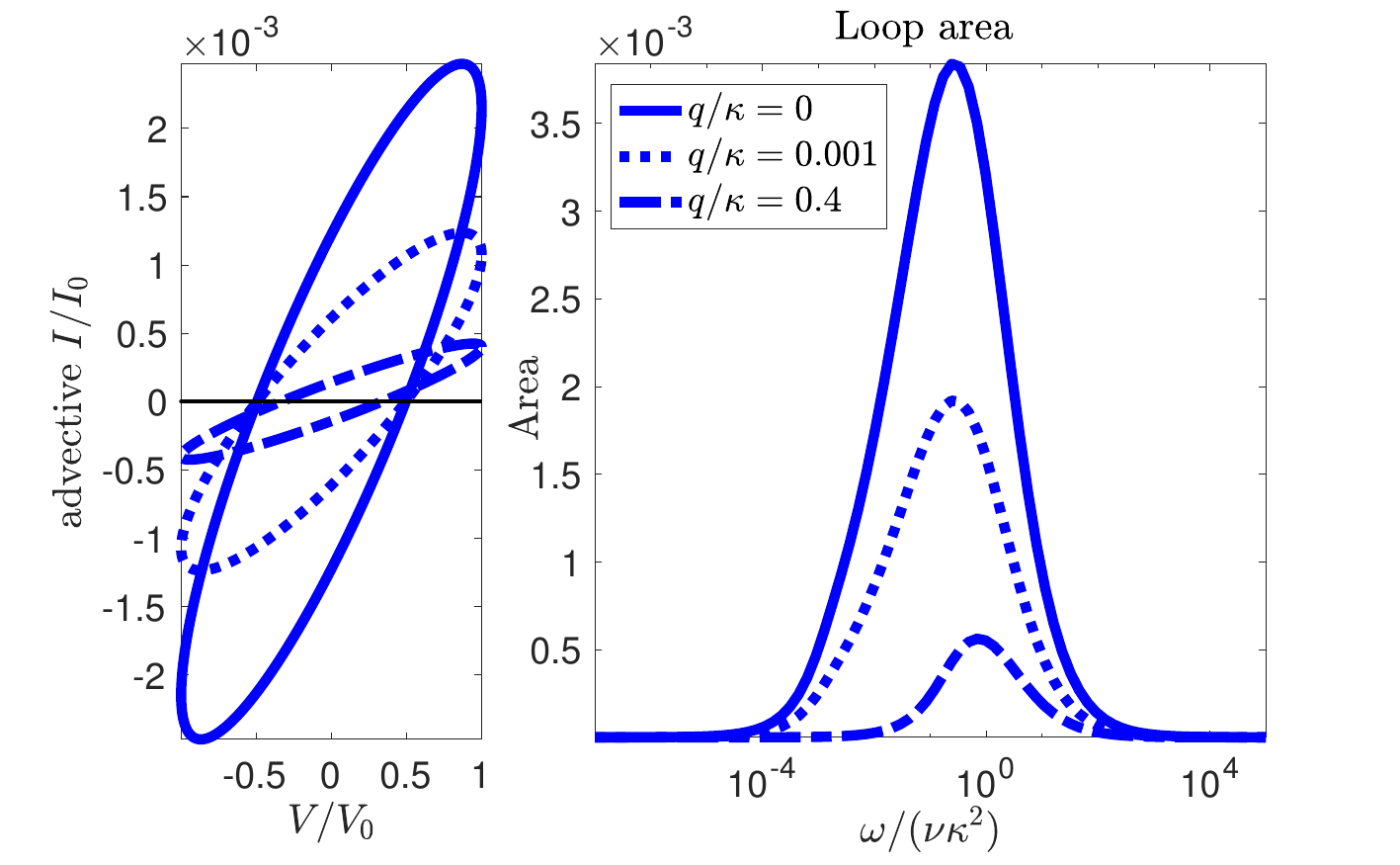}
\vspace{-0pt}
\end{center}
\caption{{Left panel: Hysteresis loops of advective current $\frac{I}{I_0} \equiv \frac{1}{I_0} \langle \rho \Re \left\{u(z,t)\right\} \rangle$ from Eq. \rr{Iss} vs.-voltage for various values of the wall charge wavevector $q$. Right panel: Area between advective current \rr{Iss} and voltage. Currents in the left panel were evaluated at frequencies that maximize the respective areas. Thus, although the average velocity over the width of the channel vanishes, the current remains nonzero and gives rise to traces of memory, see the ensuing discussion. Both panels refer to the electrolyte oscillatory channel flow \rr{vphi1} developed in section \ref{sec: par} when both walls carry the same charge distribution $\sigma^\pm = \sigma_0 \cos q x$ giving rise to the potential \rr{phi1}. }
\label{ivareaq}}
\vspace{-0pt}
\end{figure*}

\subsection{Channel memory}
\begin{figure*}
\vspace{5pt}
\begin{center}
\includegraphics[height=3in,width=6.4in]{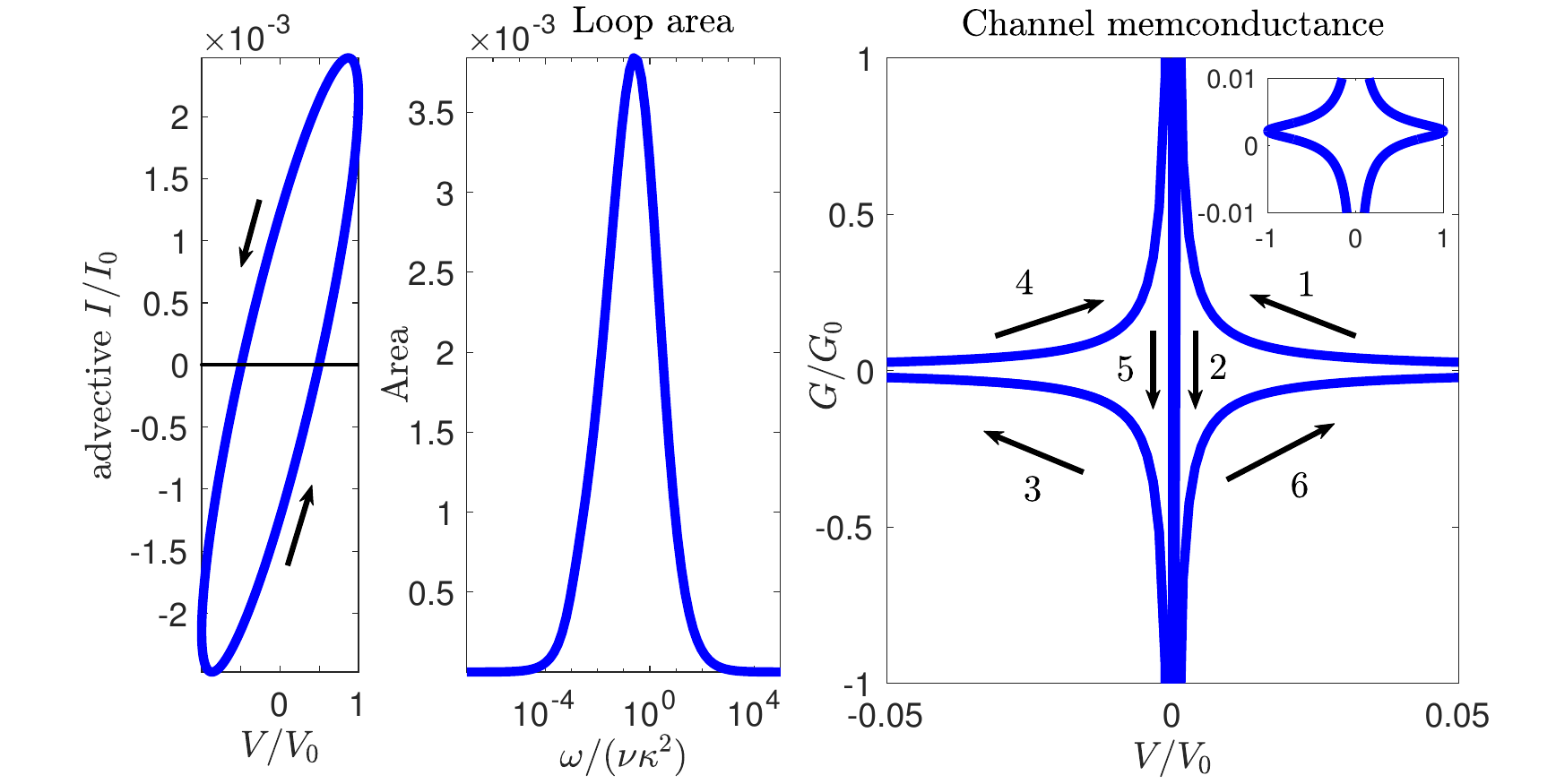}
\vspace{-0pt}
\end{center}
\caption{Channel geometry. Left panel: Advective current $\frac{I}{I_0} \equiv \frac{1}{I_0} \langle \rho \Re \left\{u(z,t)\right\} \rangle$ from Eq. \rr{Ichannel} and with the scaling $I_0$ from \rr{I0V0}, with arrows denoting the direction of voltage sweep in time, vs.-voltage hysteresis loop evaluated at 
$\omega = 0.26 \times \nu \kappa^2$ Hz which maximizes the loop area (i.e. maximum energy dissipation, cf. the area maximum in the center panel). Center panel: Area between advective current \rr{Ichannel} and voltage. 
The frequency that maximizes the area of the loop determines the characteristic `memory' time $\tau_M$ \rr{tauM}.
Right panel: Dimensionless memconductance vs. dimensionless voltage Eq. \rr{memconductance}. Numbers indicate the order at which the voltage is swept (denoted by arrows). The memconductance diverges as the voltage approaches its zero value. This figure should be compared to  \citet[Figs. 4 \& 5]{Krems2010}. Inset: same figure but for small memconductance values. {All panels refer to the electrolyte oscillatory channel flow developed in section \ref{sec: par} when both walls carry the same charge distribution $\sigma^\pm = \sigma_0 \cos q x$ giving rise to the potential \rr{phi1}. }
\label{memory_channel}}
\vspace{-0pt}
\end{figure*}

The longitudinal current $i_x$, averaged over the width of the channel becomes
\be \label{ix}
\langle i_x \rangle = \langle \rho \Re \left\{u(z,t)\right\} \rangle +  \Sigma  E \cos \omega t 
\ee
 where $u$ is given by \rr{vphi1} and in the Debye-H\"{u}ckel approximation $\Sigma=\epsilon D \kappa^2$ is the bulk conductivity.
With the scaling \rr{I0V0}
we nondimensionalize the current \rr{ix} and obtain
$
\frac{\langle i_x \rangle}{I_0} =\frac{1}{I_0} \langle \rho \Re \left\{u(z,t)\right\} \rangle + \frac{D\eta}{\epsilon\left(\frac{\sigma_0}{\epsilon \kappa} \right)^2}\cos \omega t .
$
The coefficient $ \frac{D\eta}{\epsilon\left(\frac{\sigma_0}{\epsilon \kappa} \right)^2}$ can be understood as an Osterle or Levinson number
\citep[Eq. (1.25)]{Levine1975a}, and considering standard material parameters it can be shown that it is of $O(1)$. Since this term represents an oscillation with the same phase as the driving field, it is of limited interest; we concentrate instead on the first term of the right hand side of \rr{ix} which is the advective current  $\frac{I}{I_0} \equiv \frac{1}{I_0} \langle \rho \Re \left\{u(z,t)\right\} \rangle$ in the channel with symmetric wall charge distribution, that is the electric potential is \rr{phi1}, $\rho = -\epsilon  \kappa^2 \phi$
{
and the velocity field is Eq. \rr{vphi1}. 
Simple dimensional analysis arguments show that the advective current is self-similar
\be \label{Iss}
\frac{I}{I_0}  = \mathscr{F}\left( \frac{\delta}{h}, \kappa h, q h\right)
\ee
with respect to the stated dimensionless parameters, where $\displaystyle \delta = \sqrt{\frac{2\nu}{\omega}}$ is the Stokes penetration depth defined in \rr{kdelta}. In \rr{Iss} we also averaged over one period of wall charge variation in the longitudinal direction $x$.  In the left panel of figure \ref{ivareaq} we display the advective current \rr{Iss} vs. the voltage 
$V/V_0 = \cos\omega t$ for various values of the wall charge wavenumber $q$. In the right panel of the same figure we display the area of the $I$-$V$ loops. This figure establishes what we have already stressed, that, although the average velocities over the channel width vanish, the advective current is non-zero.

The advective current \rr{Iss} is a rather long expression, which simplifies somewhat in the limit of $q\rightarrow 0$
\begin{align}
\frac{I}{I_0} =& \Re \left\{ \frac{1}{2 \left(k^{2}+\kappa^{2}\right)^{2} h \left(k h \cos \! \left(k h \right)-\sin \! \left(k h \right)\right)} \left[\left(h \left(\kappa  \,k^{2}-\kappa^{3}\right) \coth \! \left(\kappa  h \right)+h^{2} \kappa^{2} \left(k^{2}+\kappa^{2}\right) \mathrm{csch}\! \left(\kappa  h \right)^{2}-2 k^{2}\right) k \cos \! \left(k h \right) \right. \right.
\nonumber
\\
&
\left. \left.-2 \kappa  \sin \! \left(k h \right) \left(h \,k^{2} \kappa  \left(\coth^{2}\left(\kappa  h \right)\right)+\frac{\left(-3 k^{2}-\kappa^{2}\right) \coth \left(\kappa  h \right)}{2}+\frac{\mathrm{csch}\left(\kappa  h \right)^{2} h \kappa  \left(k^{2}+\kappa^{2}\right)}{2}\right)
 \right]e^{-i\omega t} \right\}
  \label{Ichannel}
\end{align}
}
which, can likewise become of the same order as the second term in \rr{ix} by a prudent choice of the parameters. 
On the left-most panel of figure \ref{memory_channel} we display the hysteresis loop formed by the advective current \rr{Ichannel} and the applied voltage $V/V_0 = \cos\omega t$. On the center panel of the same figure we display the area between the advective current \rr{Ichannel} and voltage vs. the dimensionless group $\omega / (\nu \kappa^2)$. 
The frequency that maximizes the area of the loop, determines the characteristic `memory' time $\tau_M \sim \frac{2\pi}{0.26\nu\kappa^2}$, cf. \cite[Eq.(3) \& Fig. 3C]{Robin2023} and  \cite[Fig. 3(c)]{Kamsma2023}. For standard material parameters, $\nu = 10^{-5} \; \textrm{m}^{2}\textrm{s}^{-1}$ and $\kappa = 10^7 \textrm{m}^{-1}$ we obtain the estimate 
\be\label{tauM}
\tau_M \sim 241 \mu\textrm{s}. 
\ee
We note that the above analysis can be repeated also for the antisymmetric wall charge distribution leading to the electric potential \rr{phi2} and the corresponding velocity fields.

\subsection{Channel memconductance}
Folllowing the definition of memconductance introduced by \cite{Chua1976,Pershin2011}, cf. Appendix \ref{sec: memresistive}, we 
consider expression \rr{Ichannel} (and thus for simplicity, we consider only the $q=0$ case from now on) for the advective current and form the conductance
$
G = A \frac{ I }{V},
$
which can take the somewhat simpler self-similar form 
\be\label{memconductance}
G\left( \kappa h, kh; \omega t\right) = G_0 \frac{I/I_0}{V/V_0}, 
\ee
where $G_0 = \frac{A \sigma_0^2}{\eta L}$, $A$ is the channel cross-sectional area over which the current is flowing, and we remind the reader that $k^2 = \frac{i\omega}{\nu}$, cf. Eq. \rr{kdelta}.
$G$ in \rr{memconductance} is history-dependent. 
In the left-most panel of figure \ref{memory_channel} 
we display the current-voltage hysteresis loop as the voltage is swept and in the right-most panel we display the memconductance \rr{memconductance}. It is clear that the conductance diverges as the voltage approaches its zero value and can become negative. 
This behavior should be contrasted to Figures 4 \& 5 of \citet{Krems2010} whereby the models employed \emph{and} the molecular dynamics simulations displayed in the insets therein, have exactly the behavior displayed in Fig.  \ref{memory_channel}.

{
In Appendix \ref{sec: Conductance} we argue that this behavior is generic for systems whose $I$-$V$ hysteresis loops are ellipses (as the present loops are). }

\begin{figure*}
\vspace{5pt}
\begin{center}
\includegraphics[height=3in,width=6.5in]{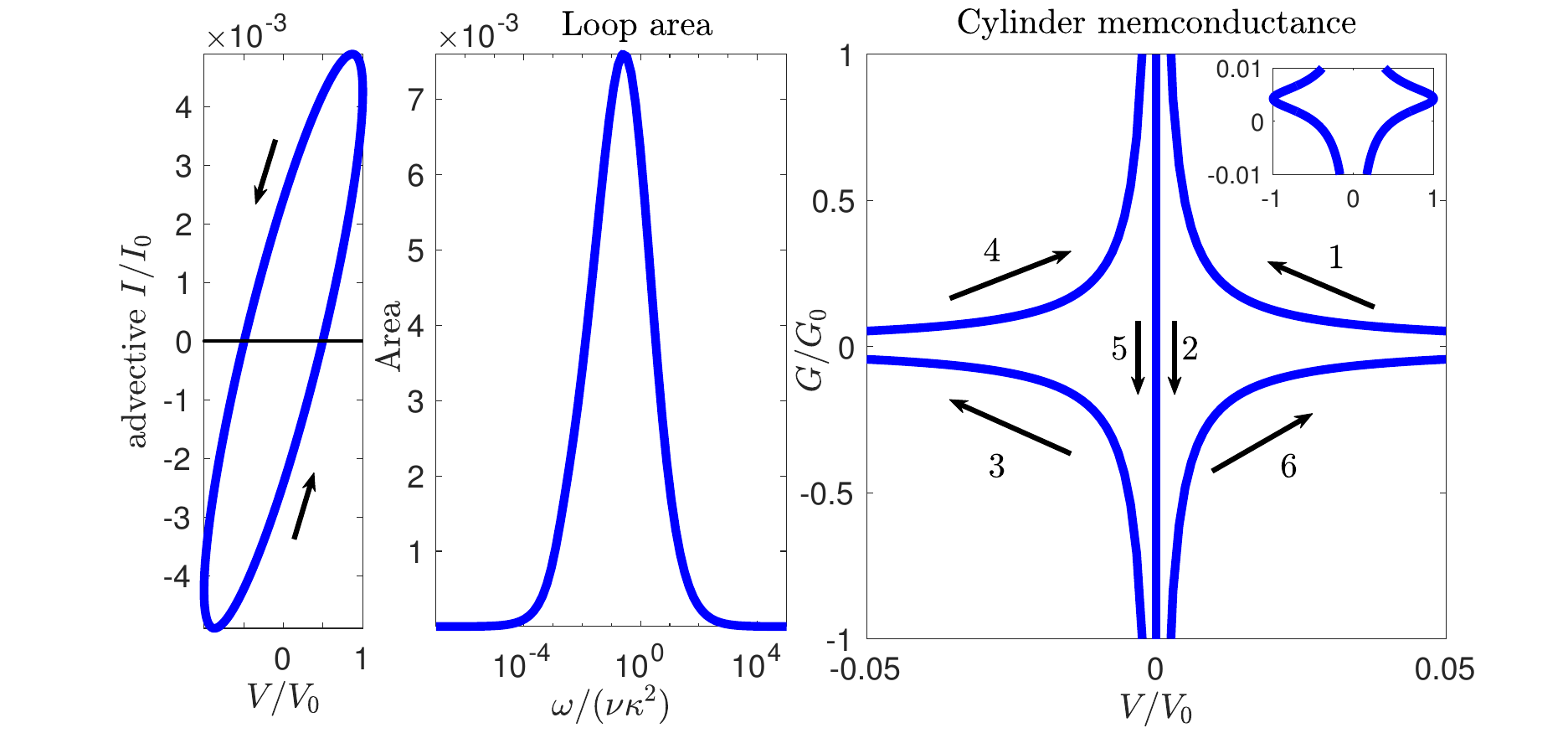}
\vspace{-0pt}
\end{center}
\caption{Cylinder geometry. Left panel: Advective current $\frac{I}{I_0} \equiv \frac{1}{I_0} \langle \rho \Re \left\{v_z(r,t)\right\} \rangle$ from Eq. \rr{Icylinder} and with the scaling $I_0$ from \rr{I0V0}, with arrows denoting the direction of voltage sweep in time, vs.-voltage hysteresis loop evaluated at 
$\omega = 0.29 \times \nu \kappa^2$ Hz which maximizes the loop area (i.e. maximum energy dissipation, cf. the area maximum in the center panel). Right panel: Area between advective current \rr{Icylinder} and voltage. 
The frequency that maximizes the area of the loop determines the characteristic `memory' time $\tau_M$. Right panel: Dimensionless memconductance vs. dimensionless voltage Eq. \rr{memconductance} (with $h\rightarrow a$). Numbers indicate the order at which the voltage is swept (denoted by arrows). The memconductance diverges as the voltage approaches its zero value. This figure should be compared to  \citet[Figs. 4 \& 5]{Krems2010}. Inset: same figure but for small memconductance values. {
All panels refer to the electrolyte oscillatory flow in a cylinder developed in section \ref{sec: cylinder} when the wall carries the same charge distribution $\sigma = \sigma_0 \cos qz$ at $r=a, $ giving rise to the potential \rr{phirz}. }
\label{memory_cylinder}}
\vspace{-0pt}
\end{figure*}

\subsection{Cylinder memory}
We repeat the above for the case of the cylinder, whose axial velocity is \rr{Stokespsi} with \rr{psir}. The advective current $I/I_0 =  \langle \rho \Re \left\{v_z(r,t)\right\} \rangle/I_0$ 
 is again self-similar
{\be \label{Isscyl}
\frac{I}{I_0}  = \mathscr{G}\left( \frac{\delta}{a}, \kappa a, q a\right)
\ee
with respect to the stated dimensionless parameters, where $\displaystyle \delta = \sqrt{\frac{2\nu}{\omega}}$ is the Stokes penetration depth defined in \rr{kdelta} and in \rr{Isscyl} we also averaged over one period of wall charge variation in the longitudinal direction $z$. The advective current \rr{Isscyl} again simplifies somewhat in the limit of $q\rightarrow 0$
\begin{align}
&\frac{I}{I_0}  =\Re \left\{  \frac{1}{I_{1}\! \left(\kappa  a \right)^{2} \left(k^{2}+\kappa^{2}\right)^{2} \left(J_{0}\! \left(k a \right) a k -2 J_{1}\! \left(k a \right)\right) a} \left[\left(2 a \,\kappa^{2} \left(k^{2}+\kappa^{2}\right) J_{1}\! \left(k a \right)-k J_{0}\! \left(k a \right) \left(a^{2} \kappa^{2} k^{2}+a^{2} \kappa^{4}+4 k^{2}\right)\right) I_{1}\! \left(\kappa  a \right)^{2}
\right.\right. \nonumber \\
& \left.\left.-2 \left(\left(-4 k^{2}-2 \kappa^{2}\right) J_{1}\! \left(k a \right)+a k \,\kappa^{2} J_{0}\! \left(k a \right)\right) \kappa  I_{0}\! \left(\kappa  a \right) I_{1}\! \left(\kappa  a \right)+a \,\kappa^{2} I_{0}\! \left(\kappa  a \right)^{2} \left(\left(-4 k^{2}-2 \kappa^{2}\right) J_{1}\! \left(k a \right)+J_{0}\! \left(k a \right) a k \left(k^{2}+\kappa^{2}\right)\right)
\right]e^{-i\omega t} \right\}. 
\label{Icylinder}
\end{align}
}
%
In the left panel of figure \ref{memory_cylinder} we display the hysteresis loop between the advective current $I/I_0 =  \langle \rho \Re \left\{v_z(r,t)\right\} \rangle/I_0$ and the applied voltage $V/V_0 = \cos\omega t$. 
In the center panel of the same figure we display the area between the advective current \rr{Icylinder} and voltage vs. the dimensionless group $\omega / (\nu \kappa^2)$. 
The frequency that maximizes the area of the loop, determines the characteristic `memory' time $\tau_M$ which is of the same order of magnitude as estimate \rr{tauM} in the channel case. 

The memconductance in the cylinder case is given again by Eq. \rr{memconductance} by effecting the change $h \rightarrow a$ and considering as cross-sectional area $A = \pi a^2$. We display in the right-most panel of figure \ref{memory_cylinder} the memconductance which diverges as the voltage approaches its zero value and can become negative. 
This behavior can again be contrasted to Figures 4 \& 5 of \citet{Krems2010}.

\section{Discussion}
Electroosmosis is an effective means of charge transport, whereby the electrolyte adjacent to a uniformly charged wall
is set into motion by an externally applied (DC) electric field. Here we showed instead that in an AC field a capillary with nonuniformly charged walls, it will give rise to vortex and other laminar flow behavior. These types of flows can be employed
for mixing and solute transport. The vortex flow patterns obtained resemble those observed in experiments with soft-lithography polydimethylsiloxane (PDMS) and glass wall channels
\cite{Stroock2000,Stroock2003}. 
In addition the non-vortex patterns shown in Appendix \ref{sec: trigh} where the wall charge distribution can change sign
(once) could lead to stronger current rectification as shown in a recent numerical comparative study
\cite{Kamsma2023}. 
We established the presence of unidirectional velocity as developed in Fig. \ref{newplot} and discussed in section \ref{sec: unidirectionalchannelperp}.

{All oscillatory systems we examined in sections \ref{sec: par} and \ref{sec: cylinder} show traces of memory.
Although the velocity field vanishes after averaging over the width of the channel or cylinder cross-section, the 
advective currents are non-zero, giving rise to $I$-$V$ hysteresis loops, which are thereby employed to develop a concept of memory time. 
The concept of memory developed in section \ref{sec: memory}, and in particular the negative and diverging conductance in the context of ionic solutions in a channel or cylindrical capillary, as displayed in the right-most panel of Fig.  \ref{memory_channel}, resembles the presence of negative and diverging capacitance in related theoretical  \cite{Krems2010} and experimental works  \cite{Wang2012}. Analogous concepts were developed in the past in the context of solid-state configurations, see for instance \cite{Martinez2010,Kirkinis2025PGE}. }

Multiple alternative configurations can likewise be studied. For instance,  
one can envision flow in the annular region between two pipes of 
radii $a$ and $b>a$ with the same or opposite wall charge distribution. Then, in addition to the Bessel functions
$J_n$ and $I_n$  employed in section \ref{sec: cylinder}, one has to consider their (diverging at the origin) counterparts $Y_n$ and $K_n$, 
respectively. Hence, the vortex pattern will resemble the left or right panel of Fig. \ref{psi12}, depending on the relative
phase of charge modulation between the inner and outer tubes.

\begin{acknowledgments}
We acknowledge the support of the NSF grant number DMR-2452280. We are grateful to an anonymous referee for suggesting the differential operator decomposition in Eq. \rr{psi4r} and for providing numerous comments that significantly improved the manuscript. 
\end{acknowledgments}

\appendix
\section{\label{sec: appendixajdari}The steady streamfunction of Ajdari \cite{Ajdari1996}}
When the electric field is static \cite{Ajdari1995,Ajdari1996}, the vorticity equation in terms of the streamfunction \rr{psi0}, acquires the form
\be
\eta \nabla^4 \psi - \epsilon \kappa^2 E_\parallel \partial_z \phi =0.
\ee
When the $x$-variation of the velocity field is proportional to $\cos q x$, the streamfunction $\psi (x,z) = \cos qx Z(z) $ 
leads to the single equation for $Z(z)$
\be
\partial_z^4 Z-2q^2\partial_z^2 Z + q^4 Z -  \frac{\epsilon \kappa^2 E_\parallel}{\eta}  \partial_z \phi=0, 
\ee
which is the counterpart of Eq. \rr{Z4} by setting $\omega = k =0$ in \rr{kdelta}. This form of solution carries 
a certain degeneracy associated with the twice-repeated wavenumbers $\pm i q$. 
As shown by Ajdari \cite[Eq.  (9) \& (15)-(19)]{Ajdari1996}, for the antisymmetric wall charge distribution giving
rise to the electric potential 
\rr{phi2}, 
 the stationary streamfunction becomes
\be \label{ajdaripsi}
\psi (x,z)=\cos qx\left[ (A_s+D_sz) \cosh qz + (B_s+C_sz)  \sinh qz + \frac{E_\parallel \sigma_0}{\eta \kappa^2\cosh Qh}  \cosh  Qz\right], 
\ee
where the subscript  $s$ denotes the integration constants of the \emph{static} problem,  
$B_s = D_s =0$, and 
\be \label{ajdariD}
\left( \begin{array}{c}
	A_s \\
	C_s
\end{array} \right) = 
 \frac{E_\parallel \sigma_0}{\eta \kappa^2\cosh Qh}
\left( \begin{array}{c}
	 \frac{\sinh \! \left(q h \right) \sinh \! \left(Q h \right) Q h - \left(\cosh \! \left(q h \right) h q +\sinh \! \left(q h \right)\right)  \cosh \! \left(Q h \right)}{q h +\sinh \! \left(q h \right) \cosh(qh)}\\
\frac{q\sinh \! \left(q h \right) \cosh \! \left(Q h \right)  -Q \cosh \! \left(q h \right) \sinh \! \left(Q h \right)}{q h +\sinh \! \left(q h \right) \cosh(qh)}
\end{array} \right).
\ee

\section{\label{sec: appendixPressure}Flow due to time-harmonic pressure gradient for channel and capillary}
We follow the solution of \cite[p.89]{Landau1987}. A harmonic in time pressure gradient inside a rectangular channel gives rise to a flow in the $x$-direction (in our notation) satisfying  the $x$-momentum
equation
\be
\partial_t u = c  e^{-i\omega t} + \nu \partial_z^2 u, 
\ee
where the units of the scalar $a$ are implied. 
The solution, satisfying no-slip boundary conditions at $z = \pm h$ is 
\be \label{uLandau}
u = \frac{ic}{\omega} e^{-i\omega t} \left[ 1 - \frac{\cos k z}{\cos kh} \right]
\ee
where $k^2\nu  = i\omega$, as considered in this article. 
Its average value is 
\be
\langle u \rangle = \frac{ic}{\omega} e^{-i\omega t} \left[ 1 - \frac{\tan kh}{kh} \right].
\ee
The adiabatic, small $\omega$ limit leads to $ \langle u \rangle\sim \frac{c h^2}{3\nu}e^{-i\omega t} $
which is of Poiseuille type.
The opposite limit of fast frequencies leads to a plug-type velocity profile 
$ \langle u \rangle\sim \frac{ic }{\omega}e^{-i\omega t} $ with a thin boundary layer adjacent to the channel walls.

For the case of an oscillating pressure gradient in a cylindrical capillary of radius $a$ we consider the solution
derived by Drazin \& Riley \cite[\S 5.1.2]{Drazin2006}. 
The velocity field in the axial direction $\mathbf{v} = v_z(r,t) \hat{\mathbf{z}}$ satisfies the 
Navier-Stokes for a liquid in a cylinder of radius $a$ under an oscillating pressure gradient has the form
\be
\partial_t v_z = c e^{-i\omega t} + \nu \left( \partial_r^2  + \frac{1}{r} \partial_r\right)v_z.
\ee
Its solution is 
\be \label{Drazinvz}
v_z(r,t) = \frac{ic}{\omega} \left( 1 - \frac{J_0(kr)}{J_0(ka)} \right)e^{-i\omega t} , 
\ee
where $k$ is given by the usual expression $\nu k^2 = i\omega$,  cf. \rr{kdelta}.

\section{\label{sec: anti}Antisymmetric charge distribution in a channel}

For the asymmetric charge distribution leading to the electric potential \rr{phi2}, the streamfunction $\psi(x,z,t) =\psi(z) e^{-i\omega t} \cos qx$ satisfying the vorticity equation \rr{vortpsi1}, leads the function $\psi(z)$ (solving Eq. \rr{Z4}) to acquire the form
\be \label{psiphi2}
\psi(z) = A \cosh \! \left(q z \right)+B \sinh \! \left(q z \right)+C \cos \! \left(K z \right)+D \sin \! \left(K z \right) + \frac{E_\parallel \sigma_0 }{\eta (k^2 + \kappa^2)} \frac{\cosh \! \left(Q z \right)}{ \cosh \! \left(Q h \right)}
\ee
with
$B = D =0$, and 
\be \label{A}
\left( \begin{array}{c}
	A \\
	C
\end{array} \right) = 
- \frac{E_\parallel \sigma_0 }{\eta (k^2 + \kappa^2)\cosh \! \left(Q h \right) } 
\left( \begin{array}{c}
	\frac{\sin \! \left(K h \right) \cosh \! \left(Q h \right) K +\cos \! \left(K h \right) \sinh \! \left(Q h \right) Q }{\sinh \! \left(q h \right) q \cos \! \left(K h \right)+\cosh \! \left(q h \right) K \sin \! \left(K h \right)}\\
\frac{ \cosh \! \left(Q h \right) \sinh \! \left(q h \right) q -\sinh \! \left(Q h \right) \cosh \! \left(q h \right) Q }{ \cosh \! \left(q h \right) K \sin \! \left(K h \right)+ \sinh \! \left(q h \right) q \cos \! \left(K h \right)}
\end{array} \right).
\ee

\section{Limiting behaviors}
Below we consider a limiting behavior that serves as reality check of our formulation. 
In the presence of DC electric field only, the above expressions for the streamfunction should reduce to the corresponding
time-independent expressions of \cite{Ajdari1995, Ajdari1996}.  We will perform this only for the unsteady streamfunction 
\rr{psiphi2} which was derived by consideration of antisymmetric channel wall charges leading to the electric potential \rr{phi2}.
Let the complex wavenumber $K$ in \rr{kdelta} be given by $K= K_1 +i K_2$ ($K_i$ are real numbers), where
\be
K_{1,2} = \frac{1}{\sqrt{2}} \sqrt{\sqrt{\left( \frac{\omega}{\nu}  \right)^2 + q^4} \mp q^2 }, 
\ee
and the $\mp$ sign corresponds to the real  ($K_1$) and imaginary ($K_2$) parts of $K$, respectively. 
First, it is clear that in the $k=\omega=0$ limit, the particular solutions in \rr{ajdaripsi} are identical. 
We thus consider  the effect of the limit $\omega \rightarrow 0$ on the combination  $ A\cosh(qz) + C\cos(Kz)$
in the unsteady streamfunction. In this limit, each one of these terms gives rise to identical leading order expressions 
proportional to $\omega^{-1}$,
but with the opposite sign, hence they cancel out. Retaining then the $O(\omega^0)$ terms only, shows that they 
recover the steady streamfunction of Ajdari \cite[Eq.  (9) \& (15)-(19)]{Ajdari1996}, denoted here by the terms $A_s \cosh(qz) + C_sz\sinh(qz)$ in \rr{ajdaripsi}, as a special case of 
our unsteady streamfunction.

\section{\label{sec: memresistive}Systems with memory}
We repeat here the definition of systems with memory \cite[Eq. (28) \& (29)]{Pershin2011}, driven by an applied voltage and resulting in a current
\be \label{IGV}
I(t) = G(x,V,t) V(t), \quad \dot x = f(x, V,t)
\ee
where $G$ is a memory conductance and $x$ is a vector of state variables.

\begin{figure*}
\vspace{5pt}
\begin{center}
\includegraphics[height=3in,width=4in]{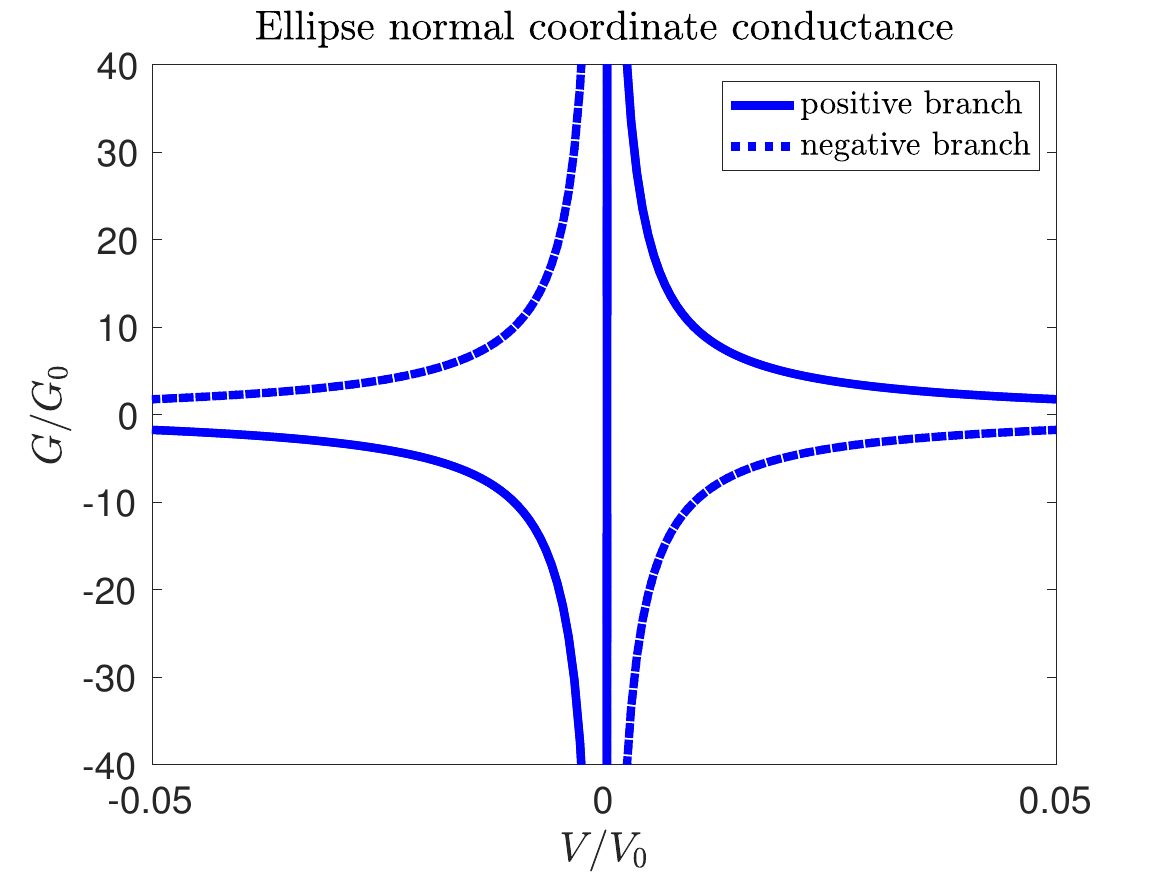}
\vspace{-0pt}
\end{center}
\caption{Conductance \rr{Ge} of an ellipse $I$-$V$ system expressed in normal coordinates. It resembles the conductance  in the right-most panel of figures \ref{memory_channel} and \ref{memory_cylinder}, as well as the memcapacitance of \cite{Krems2010} and \cite{Wang2012}. This is a generic feature of conductance (capacitance) when the current (charge) is finite at vanishing voltage. The positive (negative) branch of the ellipse corresponds to the plus (minus) sign expression in \rr{Ge}. 
\label{memory_ellipse}}
\vspace{-0pt}
\end{figure*}

{
\section{\label{sec: Conductance}Generic form of diverging conductance}
A diverging conductance is expected to be present every time the current is nonzero when the voltage is zero. Here and in \cite{Krems2010} it is clear that the current-voltage and charge-voltage loops are ellipses. This can be seen if we set the complex factor of $e^{-i\omega t}$ in \rr{Ichannel} to equal $Re^{-i\Theta}$, where $R$ and $\Theta$ are real. Thus, \rr{Ichannel} becomes
\be \label{IR}
\frac{I}{I_0} = R \cos(\omega t + \Theta). 
\ee
From elementary calculus it can now be shown that \rr{IR} along with $\displaystyle \frac{V}{V_0}  = \cos \omega t$ constitute the parametric equation of an ellipse.

For small voltages and in normal coordinates, one has $I/I_0 \sim \pm(1 - \frac{1}{2} (V/V_0)^2)$, where $I/I_0$ is the ordinate in normal coordinates and the sign denotes the branch of the ellipse. A diverging conductance at $V=0$ ensues when one expresses this current-voltage relation as a linear function of voltage: $I = GV$, whereby 
\be \label{Ge}
\frac{G}{G_0} = \pm \left( \frac{V_0}{V} - \frac{1}{2} \frac{V}{V_0} \right).
\ee
We display Eq. \rr{Ge} in figure \ref{memory_ellipse}. The resemblance to the conductance that appears in the right-most panel of figures \ref{memory_channel} and \ref{memory_cylinder} is clear. 
}

\section{\label{sec: trigh}Channel AC electroosmosis with \emph{hyperbolic} wall charge distributions}
Wall charge distributions alternative to the sinusoidal ones of section \ref{sec: equations} may also be employed. 
Apart from leading to closed-form analytical predictions, another motivation behind such choice is that they are reminiscent of those recently employed \cite{Kamsma2023}, to numerically establish
the preeminence of the commensurate current rectification in contrast to uniformly charged channel walls \cite{Kamsma2023}. 

Consider an external uniform alternating electric field $\mathbf{E} = E_\parallel e^{-i\omega t}\hat{\mathbf{x}}$ with (real) frequency $\omega$ acting on a $1:1$ electrolyte in a rectangular channel of width $2h$ (see Fig. \ref{channelAC1})
whose  upper and lower walls are insulating and inhomogeneously charged with one of the following symmetric or
antisymmetric 
distributions: $\sigma^\pm = \sigma_0 \cosh qx$, $\sigma^\pm = \pm \sigma_0 \cosh qx$, $\sigma^\pm =  \sigma_0 \sinh qx$ and $\sigma^\pm = \pm \sigma_0 \sinh qx$. Thus, there will be four distinct potentials associated with one of these
distributions satisfying Eq. \rr{helm1}, in the Debye-H\"{u}ckel approximation. 
These are displayed below and equations \rr{phiall1} and \rr{phiall2} for symmetric
and antisymmetric charge distributions on the upper and lower walls, respectively.

\begin{figure}
	\vspace{5pt}
	\begin{center}
		\includegraphics[height=4.8in,width=6.8in]{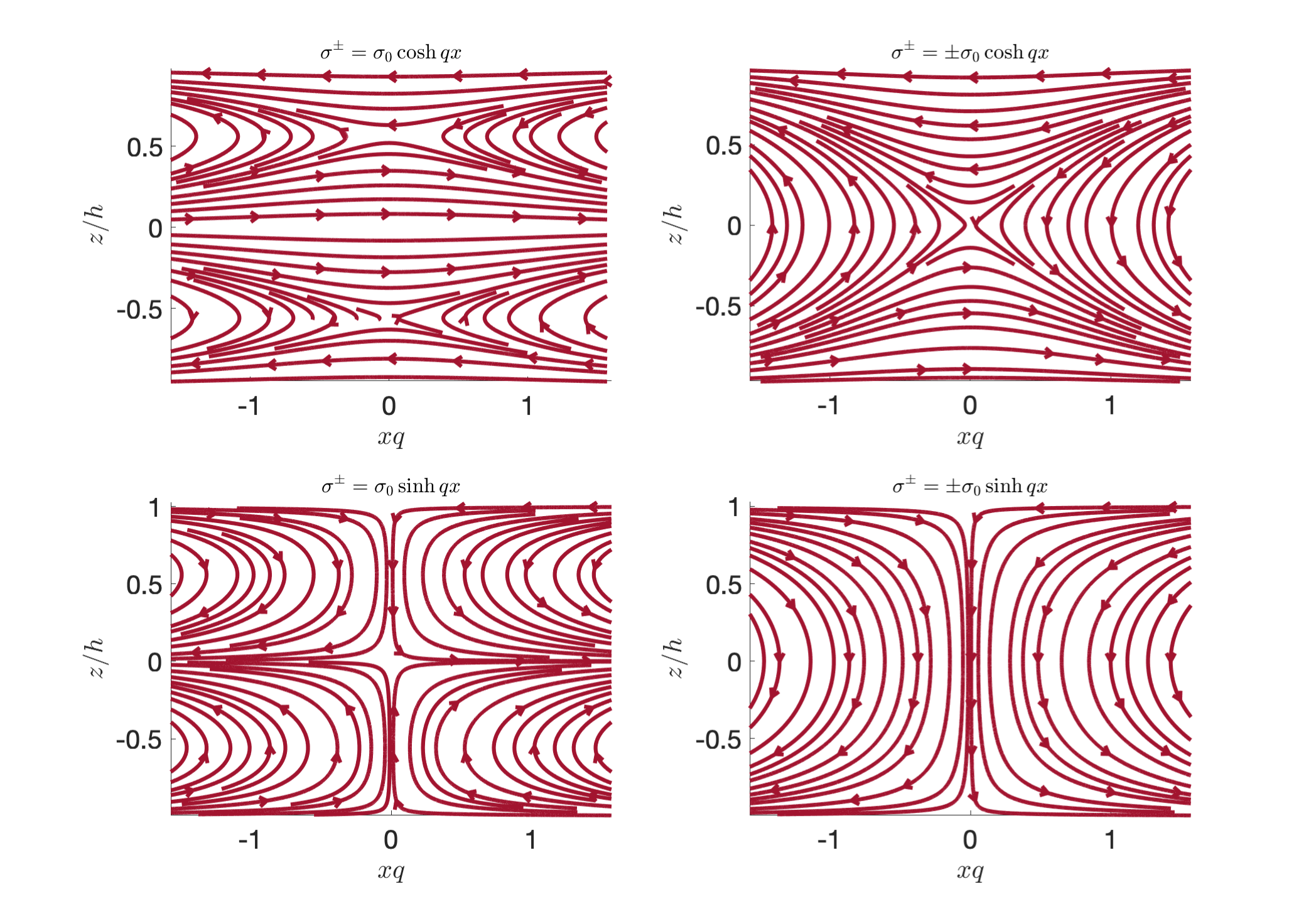}
		\vspace{-0pt}
	\end{center}
	\caption{Electrolyte (oscillating) laminar flow structure in a channel of width $h$, induced by a uniform alternating electric field
		$\mathbf{E} = E_\parallel e^{-i\omega t}\hat{\mathbf{x}}$, determined by solving \rr{Z4b}, cf. Fig. \ref{channelAC1} for the coordinate system. Here we display the dimensionless instantaneous streamline pattern determined 
		for symmetric (left column) surface charges and 
		antisymmetric (right column) surface charges along the channel walls $z=\pm  h$, corresponding 
		to the potentials \rr{phiall1} and \rr{phiall2}, respectively. 
		\label{psi12cosh}  }
	\vspace{-0pt}
\end{figure}

The vorticity equation, written in terms of $\psi$, again takes the form \rr{vortpsi1}
Assuming a streamfunction of the form
\be
\psi (x,z,t)= \psi(z) e^{-i\omega t}\cosh qx \quad \textrm{or}\quad  \psi(z) e^{-i\omega t}\sinh qx
\ee
leads to the single equation for $\psi(z)$
\be \label{Z4b}
\partial_z^4 \psi +(k^2 + 2q^2) \partial_z^2 \psi + q^2 (q^2 +k^2)\psi  - \frac{\epsilon \kappa^2 E_\parallel}{\eta}  \partial_z \phi(z)=0,
\ee
where $k$ is the complex wavenumber introduced in \rr{kdelta}, $\phi(z)$ is defined through $\phi(x,z) = \phi (z) \cos qx$
(or $\sinh qx$) and $\phi(x,z)$ is one of \rr{phiall1} or \rr{phiall2}.
The $z$ variation of $\psi(z)\sim e^{isz}$ has 
wavenumbers of the form
\be \label{s}
s  = \pm q, \quad \textrm{and} \quad s = \sqrt{ k^2 +q^2} = K_+, 
\ee
leading to a solution of the form $
\psi(z) = A \cosh \! \left(q z \right)+B \sinh \! \left(q z \right)+C \cos \! \left(K_+ z \right)+D \sin \! \left(K_+ z \right) +\psi_p(z)
$, where $\psi_p$ is the particular solution associated with the last term in \rr{Z4b} and thus determined accordingly to the form of one of the four potentials \rr{phiall1} and \rr{phiall2}. $A, B, C$ and $D$ are integration constants to be determined by the no-slip boundary condition at the channel walls (not shown). 
In Fig. \ref{psi12cosh} we display streamlines (left and right columns),
corresponding to symmetric and antisymmetric wall charge distributions, respectively.

Below we only consider the symmetric wall charges of the form $\sigma^\pm = \sigma_0 \cosh qx$ which, in the limits
$q\rightarrow 0$ and $\omega \rightarrow 0$ could be employed for unidirectional electrolyte transport. The remaining
three velocity profiles all vanish when averaged over the width of the channel. 
The motivation behind the study of these configurations lies in recent numerical experiments  \cite{Kamsma2023}, which established that current rectification is preeminent when the charge wall changes sign, in comparison to 
uniformly charged channel walls. 

Consider an external uniform alternating electric field $\mathbf{E} = E_\perp e^{-i\omega t}\hat{\mathbf{y}}$ with (real) frequency $\omega$ acting on a $1:1$ electrolyte in a rectangular channel of width $2h$ (see Fig. \ref{channelAC1})
whose  upper and lower walls are insulating and inhomogeneously charged with one of the following symmetric or
antisymmetric 
distributions: $\sigma^\pm = \sigma_0 \cosh qx$, $\sigma^\pm = \pm \sigma_0 \cosh qx$, $\sigma^\pm =  \sigma_0 \sinh qx$ and $\sigma^\pm = \pm \sigma_0 \sinh qx$. Thus, there will be four distinct potentials associated with one of these
distributions satisfying Eq. \rr{helm1}, in the Debye-H\"{u}ckel approximation, having the form
 \be \label{phiall1}
\phi (x,z)= \frac{\sigma_0}{\epsilon Q_-} \frac{\cos hQ_-z}{\sin hQ_-h} \times
\left\{ \begin{array}{l}
\cosh qx \\
\sinh qx
\end{array}
\right.
\textrm{when }\sigma^\pm = \sigma_0\times
\left\{ \begin{array}{l}
\cosh qx \\
\sinh qx
\end{array}
\right.
\ee
and
\be \label{phiall2}
\phi (x,z)= \frac{\sigma_0}{\epsilon Q_-} \frac{\sin hQ_-z}{\cos hQ_-h} \times
\left\{ \begin{array}{l}
\cosh qx \\
\sinh qx
\end{array}
\right.
\textrm{when }\sigma^\pm = \pm \sigma_0\times
\left\{ \begin{array}{l}
\cosh qx \\
\sinh qx.
\end{array}
\right.
\ee
The hyperbolic function charge distributions require a redefinition of the parameters $Q$ and $K$ which are 
\be\label{Qm}
 Q_- = \sqrt{\kappa^2 - q^2}, \quad \textrm{and} \quad K_+ = \sqrt{k^2 + q^2},
\ee
where $k$ was defined in \rr{kdelta}.

\begin{figure}
\vspace{5pt}
\begin{center}
\includegraphics[height=4in,width=6.8in]{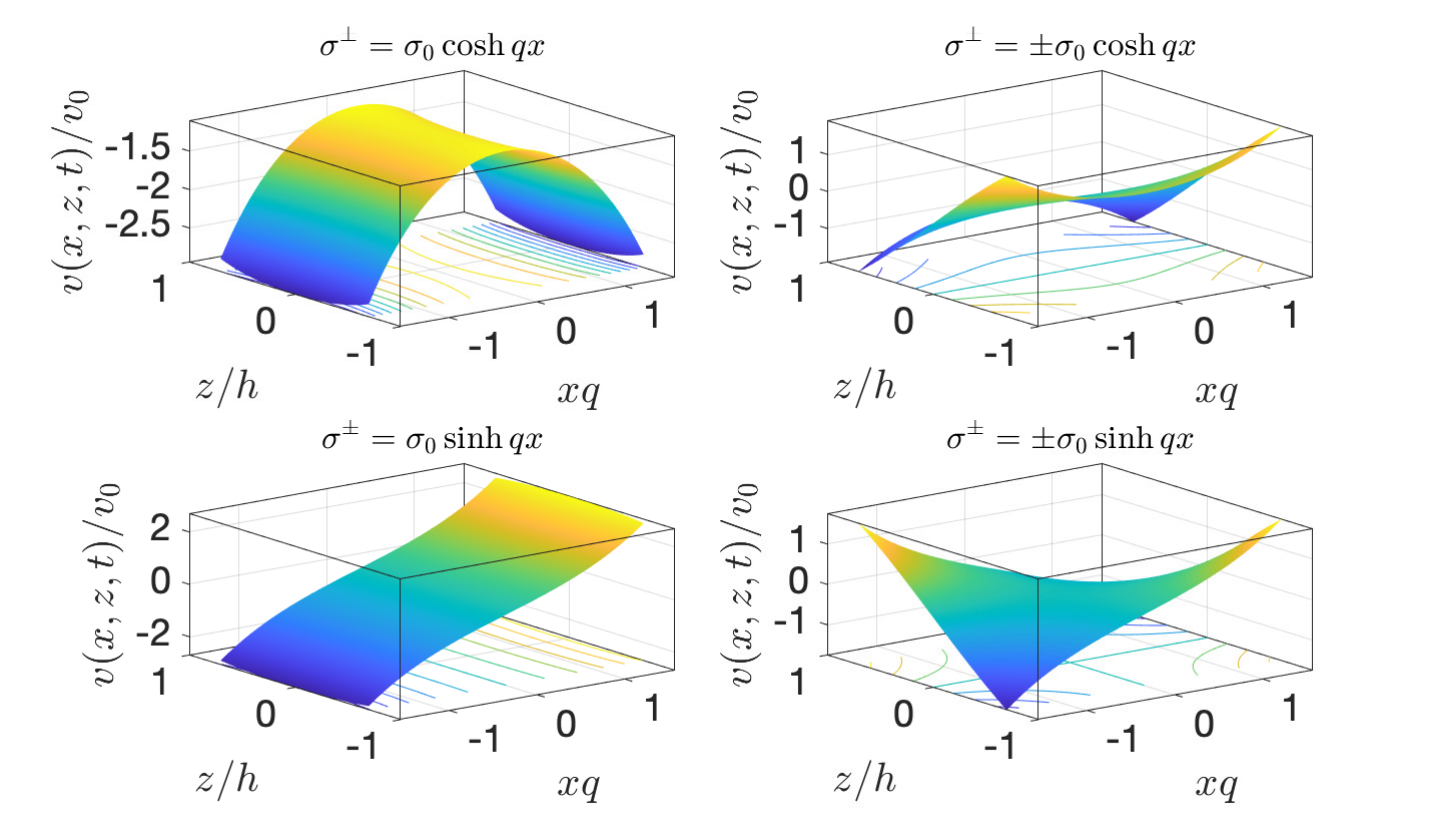}
\vspace{-0pt}
\end{center}
\caption{Electrolyte (oscillating) velocity profile $v(x,z,t)\hat{\mathbf{y}}$ scaled by the factor $v_0 = \frac{\sigma_0 E_\perp}{\eta \kappa}$, in a channel of width $h$, induced by a uniform alternating electric field
$\mathbf{E} = E_\perp e^{-i\omega t}\hat{\mathbf{y}}$ that is \emph{perpendicular} to the wall charge variation, cf. Fig. \ref{channelAC1} for the coordinate system. 
The left column displays the velocity profiles \rr{vall1} with symmetric charge distribution giving rise to corresponding
potentials \rr{phiall1}. The right column displays the velocity profiles \rr{vall2} with symmetric charge distribution giving rise to corresponding
potentials \rr{phiall2}. 
The upper left profile 
when the surface charges at the channel walls $z=\pm h$ are symmetric
$\sigma^\pm = \sigma_0 \cos q x$, can be employed for unidirectional transport of an electrolyte. 
This flow pattern is qualitatively similar to the ones obtained experimentally in \cite[figure 2]{Stroock2000} (albeit theirs being steady and expressed with respect to different boundary conditions).
\label{vcoshsinh}  }
\vspace{-0pt}
\end{figure}

Again, the velocity component $v(x,z,t)$ satisfies \rr{NS4} or 
\be \label{NS5}
\rho \partial_t v - \eta \nabla^2 {v} -\epsilon \kappa^2 \phi E_\perp e^{-i\omega t} =0. 
\ee
Assuming a velocity field of the  form $v(x,z, t) = Z(z)  e^{-i\omega t}\sinh qx$ (or $\cosh qx$), the function $Z$ satisfies 
\be
Z'' + (k^2  +  q^2) Z - \epsilon \kappa^2 E_\perp \sigma_0 \phi (z)=0, 
\ee
where $\phi(z)$ is defined as $\phi(x,z) = \phi(z) \sinh qx$ (or $\cosh qx$) and $\phi(x,z)$ is one of the four potentials
in \rr{phiall1} and \rr{phiall2}.
Thus, the velocity field acquires one of the following one forms
\be \label{vall1}
v(x,z,t) = \frac{\kappa^2 E_\perp \sigma_0e^{-i\omega t}}{\eta (k^2 + \kappa^2) Q_-\sinh Q_-h} \left( \cosh Q_-z - \frac{\cosh Q_-h}{\cos K_+ h} \cos K_+ z \right)
\times
\left\{ \begin{array}{l}
\cosh qx \\
\sinh qx
\end{array}
\right.
\textrm{when }\sigma^\pm = \sigma_0\times
\left\{ \begin{array}{l}
\cosh qx \\
\sinh qx
\end{array}
\right.
\ee
for symmetric wall charges, and 
\be \label{vall2}
v(x,z,t) = \frac{\kappa^2 E_\perp \sigma_0e^{-i\omega t}}{\eta (k^2 + \kappa^2) Q_-\cosh Q_-h} \left( \sinh Q_-z - \frac{\sinh Q_-h}{\sin K_+ h} \sin K_+ z \right) \times
\left\{ \begin{array}{l}
\cosh qx \\
\sinh qx
\end{array}
\right.
\textrm{when }\sigma^\pm = \pm \sigma_0\times
\left\{ \begin{array}{l}
\cosh qx \\
\sinh qx
\end{array}
\right.
\ee
for antisymmetric ones. 
In Fig. \ref{vcoshsinh} we display these four velocity profiles scaled by the factor \rr{v0}. 
The left column displays the velocity profiles \rr{vall1} with symmetric charge distribution giving rise to corresponding
potentials \rr{phiall1}. The right column displays the velocity profiles \rr{vall2} with symmetric charge distribution giving rise to corresponding
potentials \rr{phiall2}. 
The upper left flow pattern is qualitatively similar to the ones obtained experimentally in \cite[figure 2]{Stroock2000} (albeit theirs being steady and expressed with respect to different boundary conditions). We'll show below that it can be 
employed for unidirectional electrolyte transport.

When $\sigma^\pm = \sigma_0 \cosh qx$, the velocity profile in \rr{vall1}
averaged over the channel width leads to
\be \label{v1coshav}
\langle v \rangle = \frac{\kappa^2 E_\perp \sigma_0\cosh qx}{\eta (k^2 + \kappa^2)Q_-^2h}  \left[ 1- \frac{Q_-}{K_+ } \tan (K_+ h)\coth (Q_-h)
\right]e^{-i\omega t},
\ee
which is identical to \rr{v1av} if we replace
$\cos qx$ with $\cosh qx$ and $(Q,K)$ with $(Q_-,K_+ )$. 
Below we only consider the symmetric wall charges of the form $\sigma^\pm = \sigma_0 \cosh qx$ which, in the limits
$q\rightarrow 0$ and $\omega \rightarrow 0$ could be employed for unidirectional electrolyte transport. The remaining
three velocity profiles all vanish when averaged over the width of the channel. 
In the limit $q\rightarrow 0$ it becomes 
\be \label{uavqcosh}
\langle  v \rangle \sim \frac{\sigma_0E_\perp}{\eta h (k^2 + \kappa^2)} \left[ 1 - \frac{\kappa}{k} \tan(kh) \coth(\kappa h) \right]e^{-i\omega t}
\quad \textrm{as} \quad q \rightarrow 0.
\ee
Thus, in the long wavelength limit $q\rightarrow 0$
the behavior of the horizontal component of the velocity $u$ can also be described by the curves displayed in 
Fig. \ref{newplot} for the amplitude 
of the normalized average unidirectional velocity \rr{uavqcosh},
versus the scaled frequency $\frac{\omega}{\nu \kappa^2}$, and $v_0$ is defined in \rr{v0}. 
The adiabatic limit $\omega \rightarrow 0$ leads to same behavior expressed through the Langevin function defined in \rr{L}.

%

%

\end{document}